# Topological transmission in Suzuki phase sonic crystals

Zhen Huang[1,2], Francisco Cervera[1], Jiu Hui Wu[2]*, Martin Ibarias[1], Chongrui Liu[2], Victor M. García-Chocano[1], Fuyin Ma[2]*, José Sánchez-Dehesa[1]*

[1]Wave Phenomena Group, Universitat Politècnica de València, Camino de vera s.n. (Edificio 7F), ES-46022 Valencia, Spain

[2]School of Mechanical Engineering, Xi'an Jiaotong University, Xi'an 710049, China.

*Corresponding author: ejhwu@xjtu.edu.cn, xjmafuyin@xjtu.edu.cn, jsdehesa@upv.es

This work reports topological extraordinary properties of sound transmission through topological states in sonic crystals denominated Suzuki phase, consisting of a rectangular lattice of vacancies created in a triangular lattice. These low-symmetry crystals exhibit unique properties due to the embedded lattice of vacancies. A generalized folding method explains the band structure and the quasi-type-II Dirac point in the Suzuki phase, which is related to the underlying triangular lattice. In analogy to the acoustic valley Hall effect, the Suzuki phase contains three types of topological edge states on the four possible interfaces separating two Suzuki phase crystals with distinct topological phases. The edge states have defined symmetries with inherent directionality, which affect the topological sound transmission, and different from chirality, valley vorticity or helicity. Particularly, the existence of topological deaf bands is here reported. The propagation of topological eigenmodes on the same interface is also different, which is quantified using the acoustic Shannon entropy, making the topological transport dependent on the frequency of the edge states. Based on the abundant topological edge states of Suzuki phase crystals, a multifunctional device with acoustic diodes, multi-channel transmission, and selective acoustic transmission can be designed. Numerical simulations and measurements demonstrate the topological transmission. Our work extends the research platform of acoustic topological states to lattices with low symmetry, which opens new avenues for enriching topological states with broad engineering applications.

## I. INTRODUCTION

The discovery of the quantum Hall effect opened a new field of research in condensed matter physics, where topology describes the global behaviors of energy bands in the momentum space[1,2]. The investigation of topological phases of matter triggered fundamental discoveries, such as the quantum spin Hall effect[3,4], three-dimensional topological insulators[4,5], valley Hall insulator[6-8], crystalline insulators[9,10], and topological semimetals[11,12]. Motivated by the exotic properties of edge states, supporting transport immune to backscattering, and the development of topological phase in condensed matter systems, such topological physics concepts have been extended to classical wave systems, including photonics[13-16], acoustics[17-23], and mechanics[24-27], to realize novel topological states for wave manipulation.

For the two-dimensional (2D) topological acoustic systems, the emergence of quite a few topological edge states is associated with the existence of type-I Dirac points (DPs)[28,29], which are characterized by two fundamental features (i.e., double degeneracy and linear dispersion near the degenerate point), and guaranteed to exist at corners of the first Brillouin zone (FBZ) due to the symmetry of the lattice [28,30,31]. For example, according to group theory, the lattice with the $C_{6v}$ (or $C_{3v}$) group, which has 2D $E_1/E_2$ representation, can guarantee the presence of type-I DPs in the FBZ associated with time-reversal symmetry[21,30,32,33]. Breaking the DP can lead to the emergence of topology-related effects in acoustic systems. For example, by employing a circularly flowing into the fluid-filled region arranged in a hexagonal lattice or a honeycomb lattice [18,34], the degeneracy can be lifted without the time-reversal symmetry due to the inducing acoustic nonreciprocity[18]. The non-zero Chern number for the bands below the opened gap implies that the system supports acoustic quantum Hall effect[35]. In addition to introducing external dynamic components into the crystal, the structure of the unit cell can also be changed to break the DPs. One typical strategy is rotation-scattering mechanism to break the inversion or mirror symmetry[36,37]. The time-reversal symmetry leads to the zero Chern number, when calculates the Berry curvature over the full FBZ. However, there are strong peaks of the Berry curvature appearing near the corners of the FBZ, and gives rise to acoustic valley Hall effect[36]. Interestingly, the DPs can be superimposed[21,22]. Besides, single Dirac cone can be combined into a double cone through band folding[38], or accidental degeneracy[39]. The two degenerate modes can be hybridized to create the acoustic pseudospin states, in this way, the acoustic quantum spin Hall effect, characterized by pseudospin-dependent one-way transport of sound edge states localized at the domain walls, can be realized[39]. It is worth noting that the platforms studied above are lattices with high symmetry, such as triangular, honeycomb, Kagome, or square lattices[20,28,40]. Moreover, topologically protected edge states usually occur on one boundary, such as zigzag boundaries of triangular lattices and honeycomb lattices. Therefore, a question naturally arises for a general lattice, such as the rectangular lattice with low symmetry, whether or not there are topological edge states in the case of double degeneracy points without the limiting of type-I DPs. Moreover, what are the distinct features of the edge states if they exist?

In this work, we studied a rectangular lattice with low symmetry, known as the acoustic Suzuki phase (ASP). By employing the generalized band folding method, we explained the band structure of the Suzuki phase and discovered that it supports an accidental quasi-type-II DP in the band structure when the scatterers have a certain dimension. Drawing an analogy with the acoustic valley Hall effect, we confirmed the existence of quantized valley indices in systems with broken DPs, which can be utilized to obtain topological edge states located on the interfaces separating two systems with distinct topological phases. Numerical simulations demonstrate that both horizontal and vertical supercells with different interfaces support topological states simultaneously, with the eigenmodes of the topological states exhibiting well-defined symmetry. This unique property has implications for topological sound transmission, including the emergence of topological deaf bands. The acoustic Shannon entropy of eigenmodes indicates that even for topological states on the same type of domain wall, their degrees of localization vary, leading to changes in topological sound transmission. This wealth of topological information can be harnessed to construct versatile acoustic devices, such as acoustic diodes, multi-channel, sound source recognition, and selective sound transmitters. Additionally, simulations and experiments indicate that topological states can be employed to achieve sound energy focusing.

## II. ACOUSTIC SUZUKI PHASE AND DEGENERATE POINT

### A. Two-dimensional acoustic Suzuki phase

In the study of the physical properties of alkali halide, Suzuki found that some of them, after being doped with divalent cations, produced a new ionic compound, which has periodically distributed vacancies, and the lattice period is roughly twice the original one. On the basis of maintaining the properties of the original compound, the new compound exhibits properties specifically attributed to the lattice of vacancies[41]. Inspired by this finding, Caballero et al. introduced the sonic crystal lattice denominated Suzuki phase[42], whose scattering properties combine the properties of the underlying triangular lattice and that of the lattice of vacancies.

Figure 1a shows a tri-dimensional (3D) view of the actual two-dimensional (2D) acoustic Suzuki phase (ASP) crystal under study in this work. This crystal can be described as a rectangular lattice with lattice vectors $\boldsymbol{a}_1$ and $\boldsymbol{a}_2$, and a basis to three identical cylindrical scatterers with triangular section. Due to the absence of any rotational symmetry in the unit cell of ASP except for the identity operation[43], the ASP exhibits low symmetry. The radius of the circumscribed circle of the triangular scatterers is $R$. The central points of the four triangular prisms form a rhombus with an included angle $\gamma=60°$ and side length $a$, and the center of the rhombus coincides with the center of the cell at point O. The primitive vectors of the unit cell are $\boldsymbol{a}_1=2a\hat{\boldsymbol{x}}$, and $\boldsymbol{a}_2=a\sqrt{3}\hat{\boldsymbol{y}}$. The reciprocal space and its corresponding FBZ, which is rectangular, are plot in Fig. 1a. The four symmetry points in the FBZ are $\Gamma= (0, 0)$, $X = (\pi/2a, 0)$, $L = $ ($\pi/2a$, $\pi/(a\sqrt{3})$), and Y = (0, $\pi/(a\sqrt{3})$), which define the irreducible Brillouin zone (BZ).

Figure 1b shows the dispersion relation of the modes propagating in the ASP with parameters $a$ = 50mm, and $R$ = 0.56$a$. The acoustic band structure (see Appendix A), which is represented along the high symmetry directions of the irreducible BZ, exhibits pseudo-gaps (blue and yellow stripes) in both the ΓX and ΓY directions[42]. The formation of pseudo-gaps observed between the first and second bands is directly related to the lattice of vacancies existing in the ASP structure[42]. Since these pseudo-gaps overlap in a narrow band of frequencies, a complete band gap (red stripe) occurs at low frequencies. A nodal point exits between the 3$^{rd}$ and 4$^{th}$ bands, which is an accidental degeneracy point emerging at the low-symmetry point in reciprocal space and supported valley-Hall kink states, known as type-II Dirac point[28,44,45], as shown by the red dot (D-II) in Fig. 1b. More interestingly, the 4$^{th}$ and 5$^{th}$ bands are almost degenerate near the point X, as shown in the zoom (dashed box in Fig. 1b). In fact, we have studied the behavior of the accidental double degeneracy point (DDP) between the 4$^{th}$ and 5$^{th}$ bands as a function of the radius $R$, which represents the filling ratio in this structure. Figure 1c shows the position of the $\bm{k}$-wavevector in the direction ΓX, where the DDP appears as a function of $R/a$. For increasing values of $R/a$, the position of the DDP is located at the low-symmetry point near the X point. In this process, the frequency of the DDP decreases almost linearly with the increase of $R/a$, as shown in Fig. 1d. However, different from the type-I DP at the corners of FBZ due to the crystal symmetry, the accidental DDP formed by the 4$^{th}$ and 5$^{th}$ bands in the ASP is located at low-symmetry point in FBZ, and the cone related with the DDP lacks obvious tilted features in specific direction[28,44]. Therefore, it can be referred to as a quasi-type-II DP.

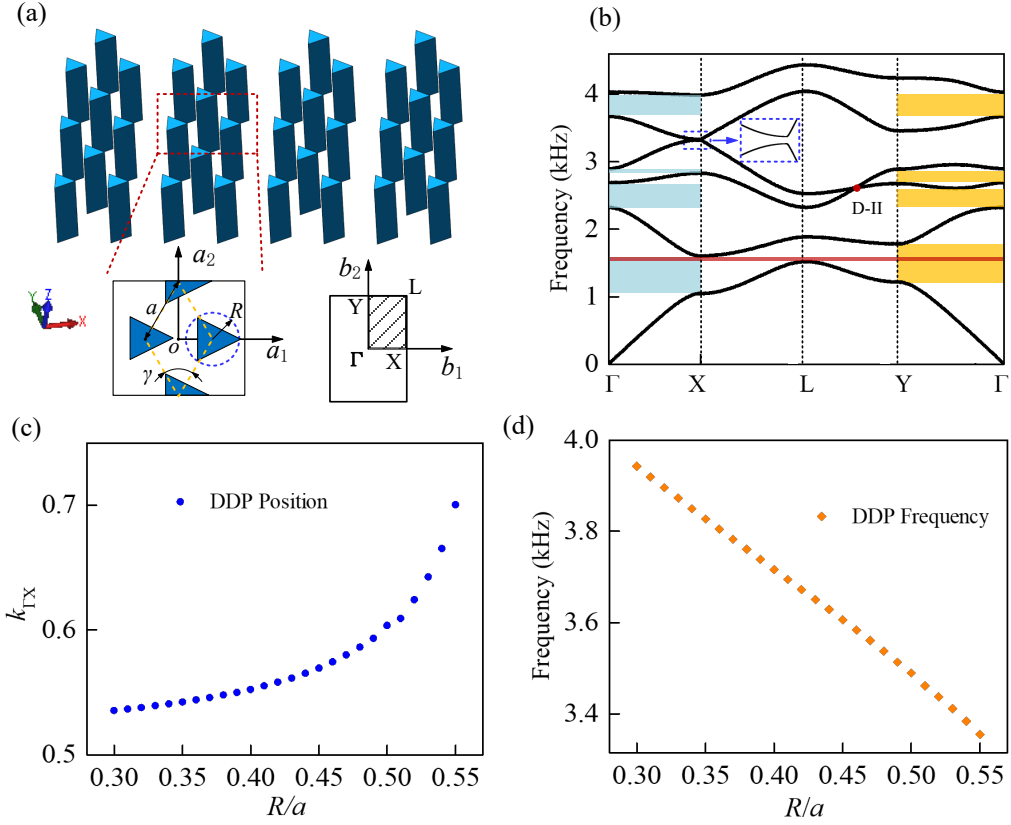

FIG. 1. (a) 3D view of Suzuki phase crystal under study. One of the two insets shows a zoom view of the 2D unit cell with the lattice vectors $a_1$ and $a_2$. The centers of the four triangular cylinders whose circumcircle radius is $R$ form a rhombus with side length $a$, and the rhombus angle $\gamma$ is 60°. In addition, the center of the rhombus coincides with the center of the Suzuki phase cell at point O. The other inset shows the FBZ of the Suzuki phase, where vectors $b_1$ and $b_2$ are the lattice vectors of the reciprocal lattice and the rectangle ΓXLY define the irreducible BZ. (b) Dispersion relation of an ASP cell. The blue shading indicates the bandgap in the ΓX direction, and the yellow area indicates the bandgap in the ΓY direction. Red shading indicates the complete bandgap. Dotted boxes indicate zoomed-in views of bands near point X. (c) The relationship between the position of DDP in $k$-space and the ratio of $R/a$. Since the length ratio of ΓX and XL is 0.75: $\sqrt{3}/2$, the relative length of ΓX is 0.75 when the relative length of XL is taken as $\sqrt{3}/2$. (d) The relationship between the frequency of DDP and the ratio of $R/a$.

### B. Generalized band folding method and degenerate point

The quasi-type-II DP in the Suzuki phase is related to the type-I DP in triangular lattice (see section I of **Supplementary Materials** for details). To reveal their relationship, we have folded the irreducible BZ of the triangular lattice into the irreducible BZ of the rectangular lattice. Figure 2a describes the folding method, which folds the discrete FBZ (blue region) of triangular lattice into the smaller FBZ (green region) of rectangular lattice. Due to the symmetry[38], the folding process can be simplified as folding region ΓPK′KΓ (including three irreducible BZs for the triangular crystal) into the region ΓXGY (containing one irreducible BZ for the ASP). And then the specific process is as follows. In the first step, considering the different geometric shape of the two areas, we have developed a method to split and classify the area to be folded into the smaller irreducible BZ. Thus, the boundary lines of the reference area (ΓXGY) are extended to intersect the large irreducible BZ at points P, Q, and M respectively. In order to obtain a simple geometric figure, draw vertical lines at K′ and M points respectively, and the intersection points are S and N respectively. Through this division, the area to be folded is divided into basic geometrical shapes. Through simple algebraic operations, $|\overrightarrow{\Gamma Y}| = |\overrightarrow{YP}| = \frac{\pi}{\sqrt{3}a}$, $|\overrightarrow{\Gamma X}| = |\overrightarrow{XN}| = \frac{\pi}{2a}$. Therefore, the yellow areas in Fig. 2a define the largest regions folded along the boundary line into the reference area and they completely cover it, as it is described in Step II. Next, considering that the triangular areas (colored in red) and the rectangular area (colored in purple) are not in contact with the boundaries of the reference area, they need to be folded into the first-level area (yellow area) first. Since $|\overrightarrow{NK}| = |\overrightarrow{SM}| = \frac{\pi}{3a}$, $|\overrightarrow{GS}| = \frac{\pi}{6a}$, the Step III in the process consists of folding the triangular and rectangular into the first-level area as shown in Fig. 2a. The process ends with Step IV in which the rectangular and triangular areas, already folded in the previous step, are folded again into the reference area, thereby finishing the folding process of the BZ of the triangular lattice into the irreducible BZ of the ASP lattice. In this process, the boundary $\overrightarrow{\Gamma X}$ of the Suzuki BZ is composed of vectors $\overrightarrow{\Gamma X}$, $\overrightarrow{NX}$, $\overrightarrow{PQ}$, $\overrightarrow{NK}$ and $\overrightarrow{K'Q}$ in the FBZ of the triangular lattice. To further demonstrating the process of band folding, we study the first band in the FBZ of the triangular lattice. Figure 2b shows the dispersion relation along the vector $\overrightarrow{PK'}$ in the boundary of FBZ of the triangular lattice. According to the geometric relationship, $|\overrightarrow{PQ}| = |\overrightarrow{\Gamma X}|$, the acoustic bands along $\overrightarrow{PQ}$ and $\overrightarrow{QK'}$ are represented by purple and cyan curves, respectively. Similarly, Fig. 2c provides the dispersion relations of $\overrightarrow{\Gamma X}$, $\overrightarrow{XN}$ and $\overrightarrow{NK}$ in the direction of the boundary vector $\overrightarrow{PK'}$, which are indicated by orange, green, and blue curves, respectively. According to the band folding procedure previously described, $\overrightarrow{\Gamma X}$, $\overrightarrow{NX}$, $\overrightarrow{PQ}$, $\overrightarrow{NK}$, and $\overrightarrow{K'Q}$ are sequentially joined into the graph shown in Fig. 2d. Orange lines in Fig. 2e show the band structure (along the direction ΓX of the FBZ) of the triangular lattice. It is compared with that of the ASP lattice (blue dotted lines) with $R=0.54a$. It is observed that the frequency of point $I_1$ in the ASP lattice has been shifted up in frequency with respect to the frequency of point $I_2$ in the triangular lattice.

Different from the band folding method usually employed in reducing BZs with similar geometrical shapes [28,46], the method developed here can be generalized to BZs with arbitrary shapes and it is more universal. In addition, from the band folding process, we infer that the quasi-type-II DP in the ASP lattice is related with the type-I DP appearing at the corners of the BZ in the triangular lattice, so it has the characteristics of an energy valley.

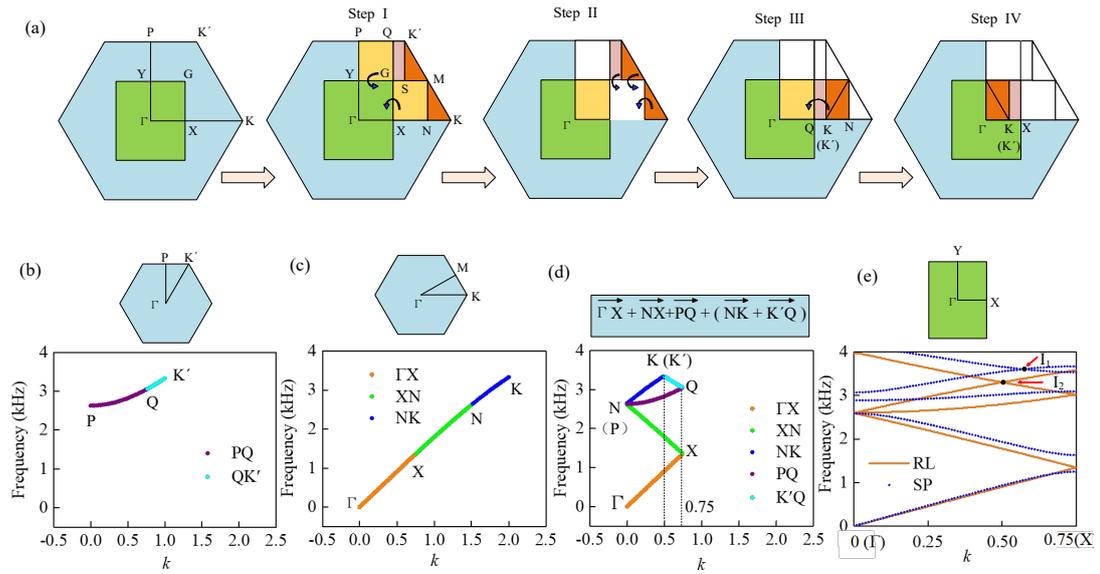

FIG. 2. (a) Sketch of the procedure showing the folding of the irreducible BZ of the triangular lattice into the one corresponding to the Suzuki phase lattice. The blue and green areas represent the FBZ of the triangular lattice and the FBZ of the Suzuki phase. The yellow areas indicate the first level of folding, the red areas indicate the second level of folding, and the purple indicates the third level. The circular arrows indicate the direction of folding. (b) Dispersion relation of triangular lattice along the boundary $\overrightarrow{PK'}$. The Q point belongs to the vector $\overrightarrow{PK'}$, and $|\overrightarrow{PQ}| = |\overrightarrow{\Gamma X}|$. (c) Dispersion relation along $\overrightarrow{\Gamma K}$ in the FBZ of the triangular lattice. Points X and N are

in the vector $\overrightarrow{\Gamma K}$, and $|\overrightarrow{\Gamma X}| = |\overrightarrow{XN}|$. (d) The first band includes modes along directions $\overrightarrow{PK'}$ and $\overrightarrow{\Gamma K}$ belonging to the area ΓPK′ and ΓMK of the triangular lattice. (e) Dispersion relation of the Suzuki phase (blue dotted line SP) along the direction ΓX (top panel) and the triangular lattice (orange line RL). $I_1$ and $I_2$ represent quasi-type-II DP in the ASP and type-I DP in the triangular lattice, respectively.

## III. TOPOLOGICAL STATES OF ACOUSTIC SUZUKI PHASE

### A. Topological phase of acoustic Suzuki phase

Part II has shown that the presence of quasi-type-II DP is secured when $R \leq 0.55a$. In addition, their frequencies decrease when the dimension of the triangular rods increases. Without loss of generality, the results reported in this section are obtained using $R = 0.54a$. Solid lines in Fig. 3a show the band structure corresponding to the ASP unit cell described in the bottom inset, where the rotation angle of rods is $\beta=30°$. The two dotted blue lines represent the 4$^{th}$ and 5$^{th}$ bands with the rotation angle of rods 0° (see the upper inset). The two bands form the quasi-type-II DP ($I_1$) as shown in the red box in Fig. 3a. The yellow stripes indicate the complete bandgaps, which are associated with the lattice of vacancies defining the ASP[42]. Particularly, Sg1 denotes the lowest bandgap. The simulations also indicate that $I_1$ occurs at the wavevector $k = 0.6695$, near the X point ($k = 0.75$) of the irreducible BZ. The rotation of the triangular rod can modify the band structure and remove the band degeneracy observed at $I_1$ [36]. Therefore, we have studied the behavior of the two modes involved in the quasi-type-II DP against a rotation period of the triangular rod, from -60° to 60°. Figure 3b shows the result obtained, showing that the mode frequencies change with the rotation angle $\beta$, undergoing a cycling process of opening, closing and opening again, which is similar to the topological transition[47]. However, the two low frequency bandgaps (orange stripes) do not undergo any bandgap closing during this rotation process. In addition, when $|\beta|=30°$, the bandgap width formed by degeneracy lifting is the largest. The SCs made of rods with rotation angles -30° and +30° are named SC-A and SC-B respectively. Figure 3c shows snapshots of the pressure patterns calculated for D1–D4 after removing the frequency degeneracy. All the patterns exhibit mirror symmetry with respect to the YZ plane. From the acoustic energy flow, which is represented by the white circular arrows, it is observed that, unlike the classical energy valley mode exchange, the modes of D1 (D2) and D4 (D3) are not exactly the same. This is caused by two main factors; one is the different lengths of the primitive vectors defining the rectangular lattice. The other is that there is an acoustic vortex at the apex of the truncated triangular rod due to the specific shape of the rod. The energy vortex is related to the shape of the truncated rod; for example, D1 and D2 in SC-A are both clockwise, while D3 and D4 in SC-B are both counterclockwise.

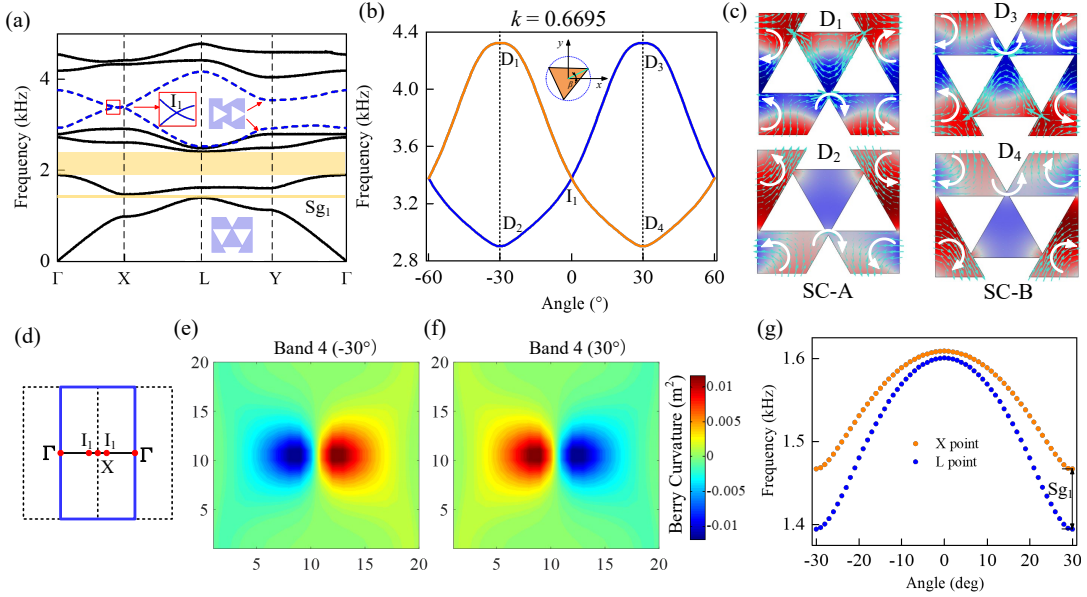

FIG. 3. (a) Band structure for the ASP lattice with $R=0.54a$. The solid lines indicate dispersion relation with $\beta=30°$ (see the inset the bottom). The dashed lines depict the fourth and fifth bands obtained with the unit cell shown in the top inset, where the triangular rods are rotated 0°. The yellow stripes indicate the complete bandgaps, where $Sg_1$ denotes the lowest frequency bandgap. (b) Dependence on the rod rotation angle ($\beta$) of the frequency of the quasi-type-II DP. The blue and orange solid lines represent the two modes involved in the DP. The inset defines the rotation angle $\beta$. (c) Pressure patterns of modes obtained when rods are rotated $\pm 30°$. Red and blue colors denote positive and negative values, respectively. The white circular arrows represent the flux of acoustic energy. (d) The FBZ of the ASP lattice. The dotted rectangles indicate two adjacent FBZs, and the solid line rectangle indicates the FBZ constructed from the center $\Gamma$ point of two adjacent FBZs. (e) and (f) show the Berry curvature distribution of the $4^{th}$ band for the rotation angles -30° and +30° respectively. (g) Rotation dependence of the frequencies at the L point of the $1^{st}$ band (blue dotted line) and at the X point of the $2^{nd}$ band (orange dotted line). The separation between both frequencies defines the bandgap width $Sg_1$.

  The system with nontrivial topological properties can be usually quantified with an invariant, a mathematical quantity that cannot change upon continuous deformations [4,25]. Since the quasi-type-II DP in the ASP is obtained by band folding of the triangular lattice, we use the Valley-Chern numbers related to the acoustic valley Hall effect to characterize the topological phases of SC-A and SC-B. We calculate the valley-Chern number of the SC in the FBZ shown in Fig. 3d, which can be obtained by integrating the Berry curvature $F_n(k)$ over the $k$-space in the 2D torus[48]. For example, for the $n$th energy band in the frequency band structure, the calculation formula of its Chern number $C_n$ can be defined as[49,50], $C_n = \frac{1}{2\pi} \iint F_n(k) d^2 k$. In order to facilitate the calculation, the torus is discretized into the same coordinate blocks, and then within each block, a same U(1) canonical transformation can be used[49,51]. Under this transformation, the distribution of the Berry curvature on the target band over the entire discrete BZ can be obtained (see section II of **Supplementary Materials** for details). Figures 3e and 3f show the Berry curvature (actually $F_n(k)/(2\pi)$) distributions on the 4$^{th}$ band of SC-A and SC-B, respectively, in momentum space, where the Berry curvature is mainly located around point $I_1$. Note that they exhibit opposite signs (represented by different color). Therefore, its integral over the full Brillouin zone is zero, while for the left half of the BZ of SC-A, the integral of the Berry curvature within $I_1$ point is -1/2. The Berry curvature integral result of SC-B is 1/2, which is the same as that of the acoustic valley state[52]. The valley-Chern number differences on both sides of the interface are quantized, $|\Delta C_4| = |C_4(SC - A) - C_4(SC - B)| = 1$. This indicates that SC-A and SC-B have different topological phases, so there will be topologically protected edge states at the interface composed of SC-A and SC-B. While there are no edge states on the interface composed of the same topological phase (see section III of **Supplementary Materials**). In addition, we calculate the frequency variation at the high symmetry points X and L in the BZ on the 1$^{st}$ and 2$^{nd}$ bands within half a rotation cycle of the rod (from -30° to 30°), as shown in Fig. 3g. The difference between both frequencies defines the bandgap width $S_{g1}$, which increases with the absolute value of the rotation angle $|\beta|$. Thus, in the ASP, the bandgap associated with the lattice of vacancies can be further widened by the rotation-rod mechanism.

For the case of acoustic valley Hall effect, spatial inversion symmetry breaking gaps between the inequivalent K and K' in FBZ gives rise to a band inversion[36]. While in Suzuki phase crystal, the quasi-type-II DP is obtained by the band folding method, and the position of DP in the momentum space is variable. Besides, in the process of band flipping of insulators supporting the acoustic valley Hall effect, a pair of eigenmodes formed by lifting degenerate point are simply flipped[53], while the case in ASP is different, due to the modulated by the scatterer geometry, this pair of eigenmodes display special symmetry. Based on these differences, the topological states obtained on the domain walls of ASP can be regarded as valleylike edge states.

**B. Superlattice band structure and topological edge states**

In order to verify the existence of topologically edge states in the ASP, we construct several supercell structures combining SC-A and SC-B. Different from the case of triangular, honeycomb, and square lattices, the interfaces of the supercell structure composed of such crystals can be divided into two types according to the distribution of SCs on both sides of the interface. That is, the cells in the supercell are arranged vertically (the number of horizontal cells is 1), and the cells in the supercell are arranged horizontally (the number of vertical cells is 1). Combined with the shape of the sonic crystals on both sides of the interface, each kind of supercell structure can be subdivided into two types. The interfaces contained in the vertical supercell are called Interface I and Interface III, while the interfaces of the horizontal supercell are named Interface II and Interface IV.

Figure 4a schematically depicts the case of a vertical supercell containing Interface I composed of SC-A (upper) and SC-B (lower), and Interface III composed of SC-A (lower) and SC-B (upper). The supercell contains 40×1 cells, of which both ends contain 10×1 cells. The zoom view of this interface (red dashed rectangle) shows that contains a rhomboid structure composed of two triangular rods. The magnified view of Interface III (red dashed rectangle) shows that Interface III contains a heteromorphic structure consisting of two trapezoidal rods. The different heterotypic structures on interfaces I and III indicate that the projected bands are also different. Floquet periodic boundary conditions are applied at the left and right boundaries of the supercell, and both ends of the supercell are set as continue boundaries.

Figure 4b shows the projected band structure of the vertical supercell. There are two kinds of edge states, the blue (ES-I) and orange (ES-III) lines representing the states on interfaces I and III, respectively, which lie within the overlapping bulk bands (black dotted points) near the degeneracy point. The quasi-type-II DP in the ASP is analogous to the valley edge state since it originates from the Dirac point of the triangular lattice, which sustains time-reversal symmetry. Therefore, the band structure is symmetric with respect to $k_x$=1. In addition, as shown in Fig. 4b, there are forward-moving states, with group velocity d$\omega$/d$k$>0, and backward-moving states with group velocity d$\omega$/d$k$<0[54]. Due to the strong perturbation of the sound field by the irregular structure on Interface III, a narrow bandgap occurs between the forward and backward propagating edge states. However, there are three bands composed of edge states at Interface I instead of the usual two. To the best of our knowledge, this phenomenon was not discussed nor reported in the existing literature on acoustic topological states.

Figure 4c shows the sound pressure patterns and sound field intensity distribution of eigenmodes $A_I$, $B_I$, and $C_I$ at Interface I, and $A_{III}$, $B_{III}$ at the Interface III calculated at $k_x$ = 0.5. The patterns of eigenmodes $A_I$, $B_I$, and $C_I$ show localization on interface I, and $A_{III}$ and $B_{III}$ are located at Interface III (zoom views are shown in the red dotted rectangle) and correspond to topologically protected edge states. In addition, the intensity distribution, indicated by the cyan arrows, shows the direction of motion of the edge states corresponding to the forward- and backward-moving states. However, as shown in Fig. 4c, the group velocity of $B_{III}$ is almost zero (flat band) and its direction of propagation is not obvious.

Figure 4d depicts the horizontal supercell containing interfaces II and IV, the same size as a vertical supercell. The projected band structure of the horizontal supercell is plotted in Fig. 4e, where the blue lines (ES-II) and red dashed lines (ES-IV) within the bandgap of the bulk, indicate the edge states on interface II and IV, respectively. At $k_y = 0.5$, we select four edge states denominated as $A_{II}$, $B_{II}$, $A_{IV}$, and $B_{IV}$, in which the subscripts II and IV represent interfaces II and IV, respectively. Figure 4f shows the distribution of acoustic pressure localized near Interface II and Interface IV. The white circular arrow indicates the propagation direction of the acoustic energy flow. Since the two SCs do not form any new geometric shape at the interfaces, the pressure pattern at the interface is directly related to that of the eigenmodes in the unit cell. For example, the $A_{II}$ mode in Fig. 4f can be obtained by just joining the patterns of the modes $D_2$ and $D_4$ represented in Fig. 3c. Similarly, the $B_{II}$ mode in Fig. 4f can also be obtained by sequentially joint the two higher frequency modes represented in Fig. 3c. For the $A_{IV}$ and $B_{IV}$ modes, the cases are the same with $A_{II}$ and $B_{II}$. In fact, this is the reason that the band structure for interfaces II and IV are the same. Within the bulk band, there is also a band (represented by gray lines pointed by the red arrows in Fig. 4e) composed of edge states near 2.6 kHz. In fact, the projected bands are composed of tilted type-II valley-Hall kink states[28,44], due to the type-II DP (D-II) in Fig. 1b.

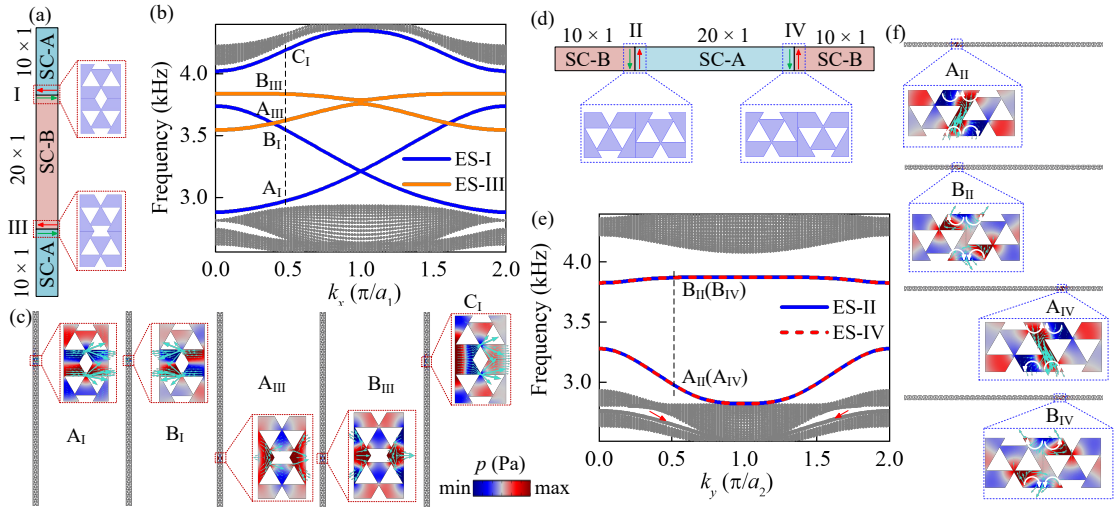

FIG. 4. (a) Scheme of the superlattice composed of SC-A and SC-B with Interface I and Interface III. The red-dotted rectangular box deploys a zoomed view of interfaces I and III. (b) Acoustic band structure of the supercell (a) around the quasi-type-II DP. The black dots represent bulk bands. The blue and orange lines represent the edge states located on interfaces I and III, respectively. (c) Acoustic pressure patterns and energy flow of eigenmodes $A_I$, $B_I$, and $C_I$ at Interface I, and $A_{III}$ and $B_{III}$ at Interface III at $k_x = 0.5$. (d) Scheme of the superlattices obtained by combining SC-A and SC-B. Their respective interfaces are denominated as Interface II and Interface IV. The blue dotted rectangles show zoom views of the two interfaces. (e) Band structure the supercell containing interfaces II and IV, where the dotted lines represent bulk modes. The blue lines (ES-II) and red dashed lines (ES-IV) represent the edge states located at Interface II and Interface IV, respectively. (f) Pressure patterns and energy flow distribution of the eigenstates denominated $A_{II}$, $B_{II}$, $A_{IV}$, and $B_{IV}$. The colored maps represent the acoustic pressure (real part) and the cyan arrows represent the energy flux.

### C. Directionality of topological edge states

The pressure patterns of edge states at the four interfaces are confined near the interface, and they have recognizable symmetries, especially at interfaces I and III. In order to understand the influence of symmetry on the transport properties of sound waves, we study the structure shown in Fig. 5a, where the four quadrants are made of different unit cells. A Cartesian coordinate system is added at the center of this structure, which is divided into four equal parts 1-4, in which the upper region (the part composed of regions "1" and "2") and the lower region (the part composed of regions "3" and "4") respectively represent different SCs. For the convenience of description, the bands ES-I, ES-II, ES-III, and ES-IV composed of edge states are divided. ES-I, for example, is divided into ES-I-1 and ES-I-2, where the frequency in ES-I-1 is less than the one in ES-I-2. Besides, we define a symmetry rule as follows; when the mode pressure map is symmetric about the coordinate axis, it is marked as "1", and when it is anti-symmetric about the coordinate axis, it is marked as "0". Therefore, two digits can express the symmetry of the SC structure. The first digit indicates symmetry about the $x$-axis, while the second digit indicates symmetry about the $y$-axis. First, let us analyze the case where the upper region is SC-B, as shown in Fig. 5b. On the bands of edge states (indicated by blue and orange solid lines in the figure), we select four edge modes (at $k_x = 0.5$ and $k_x = 1.5$): $A_1$, $B_1$, $A_1´$, and $B_1´$. Their pressure maps have the same symmetry, indicating that all have symmetry "11", enlightened in yellow.

When the upper region is changed to SC-A, the symmetry becomes complicated due to the number of bands containing edge states. As shown in Fig. 5c, the upper panels depict the high-frequency boundary states E and E´, obtained at wave vectors $k_x = 0.1$ and 0.9, respectively. Their pressure patterns indicate that their symmetry is also "11". While for edge states with lower frequency, as shown in the middle and lower panels of Fig. 5c, calculated at $k_x = 0.5$ (1.5), the symmetry of the modes $A_2$ ($A_2´$) and $B_2$ ($B_2´$) about the x-axis becomes antisymmetric and they should be labeled with "0". Therefore, the symmetries of $A_2$ ($A_2´$) and $B_2$ ($B_2´$) can be labeled "01" ("01") and "00" ("00"), respectively. Figs. 5d and 5e represent the two kinds of horizontal structures, where the patterns cannot be labeled anymore with integers due to their lack of symmetry about the x- or y-axis. However, they exhibit obvious directivity that could be defined by a fractional number.

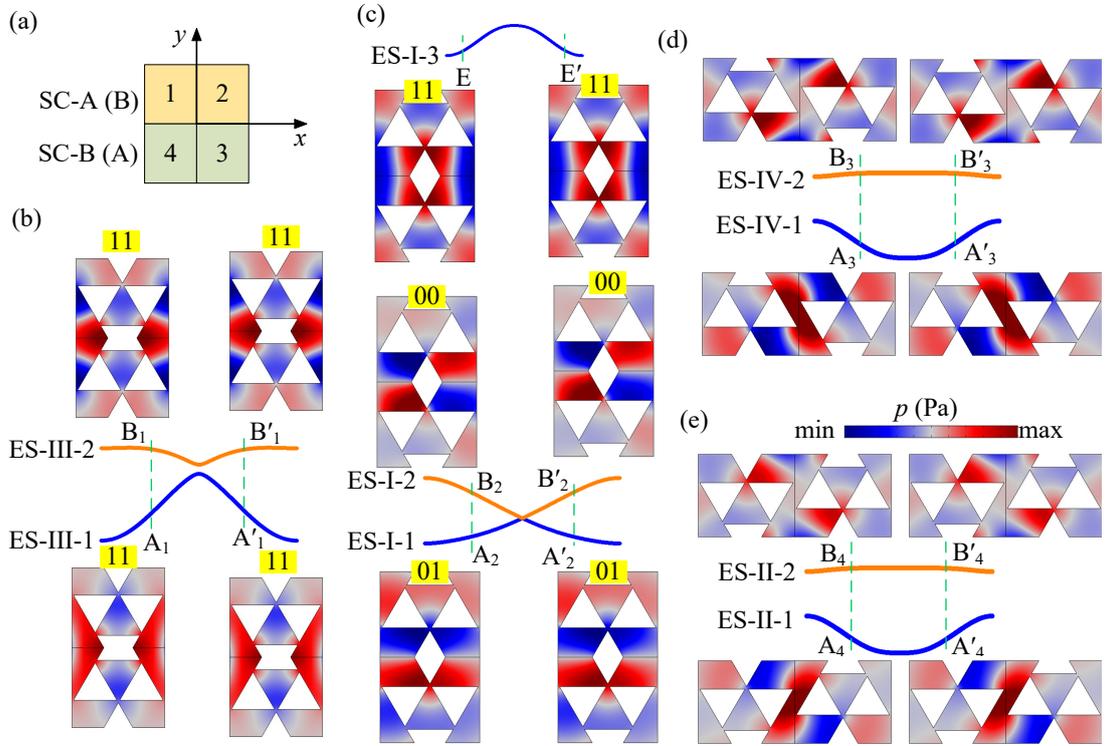

FIG. 5. (a) Scheme of the structure containing SC-A and SC-B. The SC is divided into four regions, 1 to 4, where four different interfaces can be formed. (b) Snapshots of the sound pressure pattern of the edge states located at Interface III (calculated at $k_x = 0.5$ and 1.5, respectively). The numbers enlightened in yellow represent the mirror symmetry about the xz-plane and yz-plane, respectively, "1" means symmetric, and "0" means anti-symmetric. The blue and orange lines provide the dispersion relation of the edge modes. (c) Snapshots of the sound pressure pattern of the edge state located along Interface I. The upper panel corresponds to eigenmodes at $k_x = 0.1$ and 1.9, and the middle and lower panels represent eigenmodes at $k_x = 0.5$ and 1.5. (d) and (e) depict the sound pressure patterns of edge states at interfaces IV and II

respectively. The color scale represents the magnitude of the magnitude of the real component of the pressure, going from negative values (blue) to positive values (red).

The realization of topological edge states in the acoustic Suzuki phase requires two key factors, the double degenerate point and the crystals on both sides of the domain wall with equal magnitudes of Berry curvature but opposite signs, which are also the prerequisites for valley edge sates[36,55,56]. And this is the main reason that the topological edge states in the ASP are regarded as valley-like edge states. However, different from the previous valley edge states, in the ASP, the degenerate point is not caused by lattice symmetry[37,56], but by the mechanism of generalized band folding, and it appears at low-symmetry point in the FBZ instead of the corners[36,55,56]. The projected band structure of the vertical and horizontal supercells shows that the topological edge states can appear on four boundaries of unit cell, and the frequencies and number of edge states are different because of geometric configurations at the domain walls. Due to the modulation of the scatterer shape, the edge states stem from the quasi-type-II DP in ASP display obvious symmetry about directional axis, such as *x*-axis, *y*-axis or one specified axis, which are different from the chiral edge states of the acoustic quantum Hall effect[35,57], the valley-vortex edge states of the acoustic valley Hall effect[36,37], and the helical edge states in acoustic quantum spin Hall effect[38,39]. As a consequence, the excitation of the edge modes strongly depends of the symmetry of the wave employed as the excitation source, as it was observed for acoustic modes in typical SC[58]. From the pressure pattern of the modes depicted in Fig. 5, the edge states with the lowest frequencies at Interface I ($A_2$, $A_2'$, $B_2$, and $B_2'$) are particularly interesting, since they are antisymmetric with respect to the *xz*-plane. Therefore, for an incident plane wave traveling along the *x*-axis, these edge states cannot be excited by such excitation source[58]. These edge states are said to belong to a deaf band of edge states, a phenomenon that is reported here for the first time in relation to acoustic topological states. As for eigenstates in other interfaces, they can be excited by an incident plane wave along the *x*-axis because their pressure pattern has mirror symmetry with respect to the *xz*-plane. For a comprehensive study regarding the conditions, in terms of symmetries, for the excitation of edge states by external excitation sources, see section IV in the **Supplementary Materials**.

### D. Acoustic Shannon entropy of topological edge states

The acoustic Shannon entropy (ASE) is a useful tool to characterize the spreading of acoustic eigenmodes. It was introduced by Sánchez-Dehesa and Arias-Gonzalez[56] as $S_u = -\int P(\boldsymbol{r})\ln P(\boldsymbol{r})\mathrm{d}\boldsymbol{r}$, where $P(\boldsymbol{r})$ represents a false probability distribution function defined as $P(\boldsymbol{r}) = A|u(\boldsymbol{r})|^2 / \int |u(\boldsymbol{r})|^2 d\boldsymbol{r}$, with $|u(\boldsymbol{r})|^2$ being the square norm of the total acoustic pressure and $A$ is the area of the integration domain. The ASE is then a quantity providing an information measure of the spatial localization of a given acoustic mode, and it increases with increasing uncertainty (i.e., spreading of the mode).

Figure 6 shows the calculated ASE corresponding to the band ES-I, ES-II, ES-III and ES-IV. In Fig. 6a, green, blue and orange lines represent topologically edge states ES-I-1, ES-I-2, and ES-I-3, respectively. The localization degree of the edge states is different and depends on the $k$ wavenumber. For example, the spreading of edge states belonging to ES-I-2 is smaller than the other two. Figure 6b shows the ASE distribution of the eigenmodes of the supercell including Interface III, in which the localization degree of topological states on ES-III-1 (green line) is higher (with a lower value of ASE) than that on ES-III-2 (blue line). The eigenmodes in ES-II(IV)-1 (green line) and ES-II(IV)-2 (blue line), in which the two lines are crossing, and the ASE for ES-II(IV)-2 has a slight change in the interval $0.4 < k < 1.6$, as shown in Fig.6c. The greater the Shannon entropy is, the larger the spreading of the acoustic mode is[59], hence, the corresponding propagation is different. The one-dimensional projection spaces of the vertical and horizontal supercells are shown in Fig. 6d, and the projection vector space of the vertical and horizontal supercells are $k_x$ and $k_y$ directions, respectively. According the projected band structure and ASE of edge states, when additional structures appear on the interface, the topological states are not only related to the interface of the supercell, but also related to the type of supercell (the direction of the projected space).

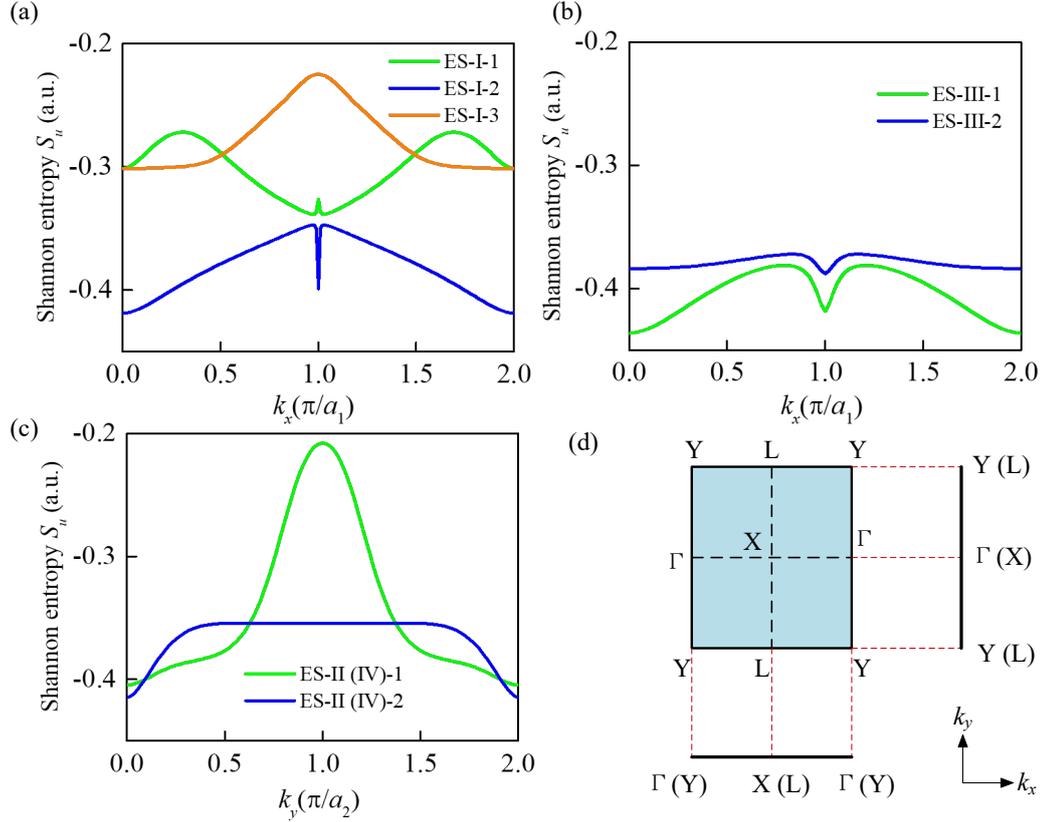

FIG. 6. (a) Acoustic Shannon entropy of eigenmodes localized at Interface I. Green, blue and orange lines represent topologically protected edge states ES-I-1, ES-I-2, and ES-I-3, respectively. (b) Acoustic Shannon entropy corresponding to eigenmodes localized at Interface III. Green and blue lines represent the edge states ES-III-1 and ES-III-2, respectively. (c) Acoustic Shannon entropy corresponding to eigenmodes localized at Interface II(IV). Green and blue lines represent the edge states ES-II(IV)-1 and ES-II(IV)-2, respectively (d) 1D projection of the vector of the 2D first BZ of the Suzuki phase in the $k_x$ and $k_y$ directions, respectively.

## D. Transport along domain walls

The paramount property of topological non-trivial systems is the exhibition of unidirectional and robust propagation with edge states existing at the interfaces between distinct topological zones. As described above, the symmetry of the pressure patterns of the edge states enriches the transmission features through topological edge modes. On the other hand, some asymmetric scattering, producing an effective component perpendicular to the incident wave vector, can easily modify the incident exciting plane wave. Therefore, when 2D plane waves impinge the structure through the four different interfaces, a variety of behaviors in transmission appear depending on the incident directions. Figure 7a shows a diagram that condenses the information regarding the bandwidths of topological edge states in vertical and horizontal supercells. In this diagram, the blue and red stripes in the first row represent the bands containing edge modes localized on Interface I (schematically depicted at the right-hand side), while the edge modes localized on Interface III (schematically depicted at the right-hand side) are represented by the green and orange stripes in the second row. Finally, cyan and purple stripes in the last row define the frequency range containing edge modes along interfaces II and IV (Interface IV are schematically depicted on the right-hand side). Since some bands have overlapping frequency regions, the edge states in these regions can perform multi-channel wave transmission, a property with interesting engineering applications. In order to study the topological transmission, the diagram is further divided into seven frequency bands, named EG-A, EG-B, EG-C, EG-D, EG-E, EG-F, EG-G, as shown in Fig. 7a. The values define the frequencies of the seven bands. It should be noted that the frequencies reported here correspond to states in the bulk bandgap, which is slightly different from that of the total dispersion relation.

Figure 7b shows the scheme of the composite SC structure employed to study the different possibilities of topological wave transport. They can be obtained because of the frequency and symmetry features of topological edge modes. The structure composed by two SC-A (blue areas) and two SC-B (orange areas) is made of 16×16 cells and contains the four interfaces I-IV. Red thick lines, with length *a*, represent the incident planes whose center is at the interfaces. In the numerical simulations, a perfect matching layer is applied around the boundaries of the structure. In addition, we introduce a defect cylinder (see the black dot around the center of the structure) that can be ignored due to the topologically, but it can create additional modes (see section 4 in **Supplementary material**). Next, we apply plane waves impinging the different ports to analyze the transmission characteristics in the combined structure, with frequencies within the overlapping bands EG-A, EG-C, EG-E, and the single band EG-F (see section V in **Supplementary Materials**).

Due to the directionality of the edge states and differentiation in ASP, sound transmission through topological states in Suzuki phase sonic crystals displays extraordinary properties, as shown in Fig. 7c. For example, when the plane wave with direction [0,1,0] at 3500Hz impinges Port-I, sound waves can only be localized on the incident Interface I, and this feature can be used as identification of the exiting source. In the EG-C band, when the plane wave with direction [1,0,0] at 3685Hz impinges Port-I, the transmission is forbidden because the symmetry of the edge states is orthogonal to that of the exiting wave. However, when the plane wave with the same frequency in direction [-1,0,0] impinges Port-III, the edge modes along interfaces III and II can be excited because they have the same symmetry. This property can be exploited to realize acoustic transmission with non-reciprocal properties, namely acoustic diodes. In addition, the multi-channel wave transport is also possible, i.e., when the waves enter in one port and it is transmitted to more than one port at the output. However, based the Shannon entropy, let us point out that the sound energy transmitted along the multichannel is not equally distributed, so that the sound energy in the other exit port could be almost zero, as shown in Fig. 7c with wave directions [0, 1, 0] at 3095Hz, 3185Hz, 3830Hz, 3838Hz and 3852Hz from Port-II. In summary, the transmission properties shows that ASP crystals can be employed to develop acoustic devices like acoustic diodes, multi-channel and selective acoustic transmission, sound source recognition, and so on.

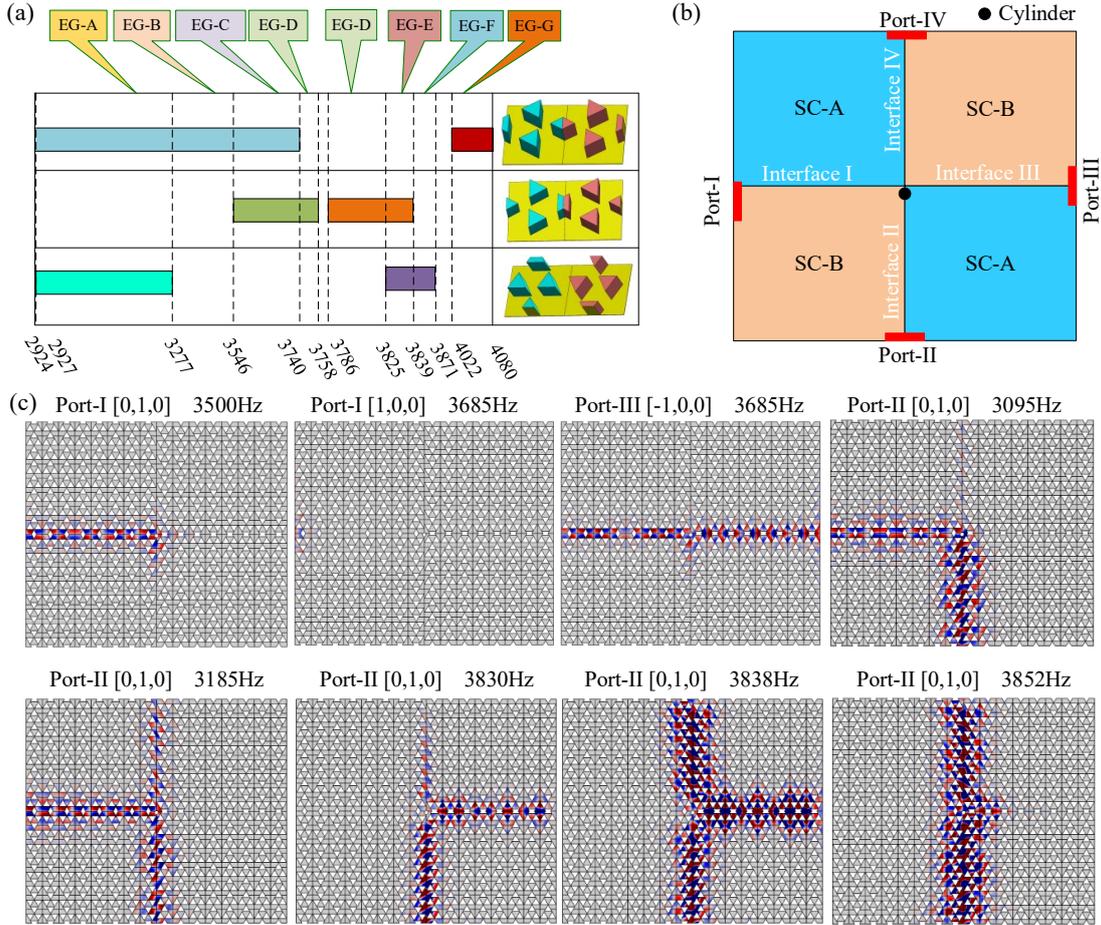

FIG. 7. Frequency spectra of the topologically protected edge states. The 1st row indicates the frequencies of states (blue and red strips) of supercell containing interface I. The 2nd row defines frequencies (green and orange strips) with states localized at interface III. The cyan and magenta strips in the 3rd row indicate the frequencies of edge states localized at interface II or IV. The bands are further divided into seven smaller frequency bands (defined on top) according to overlapping features. (b) Scheme of a SC structure made of SC-A and SC-B containing the four interfaces. The distinct edge states are excited through ports named I, II, III, and IV. The black dot indicates the auxiliary cylinder employed to disturb the field of the impinging waves arriving at the ports. (c) Propagation of a plane-wave with direction [0, 1, 0] at 3500Hz from Port-I, with direction [1, 0, 0] at 3685Hz from Port-I, with direction [-1, 0, 0] at 3685Hz from Port-III, with directions [0, 1, 0] at 3095Hz from Port-II, with directions [0, 1, 0] at 3185Hz from Port-II, with directions [0, 1, 0] at 3830 Hz from Port-II, with directions [0, 1, 0] at 3838 Hz from Port-II, and with directions [0, 1, 0] at 3852 Hz from Port-II, respectively.

## IV. EXPERIMENT OF TOPOLOGICAL STATES TRANSPORT

For horizontal supercells, the topological eigenmodes located in the bulk bandgap do not have mirror symmetry about the *XZ*-plane or *YZ*-plane (see Fig. 5) and, therefore, they are insensitive to the direction of the incident wave. In addition, compared with the vertical supercells, the values of the ASE of the horizontal supercells are larger, meaning a wider spreading of their eigenmodes. Therefore, to study the propagation properties under a given incident direction, we consider the experimental characterization of the horizontal supercell, and assume that similar results would be obtained for the vertical supercells.

Figure 8a shows the experiment setup (see **Appendix B**), in which the size of the sample made of SC-A and SC-B containing 6×6 cells. As a comparison, we designed a structure with the same size for simulation calculation. To quantify the amount of sound transmission we use the transmission loss (TL), which is defined as TL = $10\log(P_{in}/P_{out})$. Here, $P_{in}$ and $P_{out}$ represent the acoustic energy density at the entrance and exit of the sound field, respectively. The sound energy density can be calculated by the following formula (see **Appendix B**), $P_{in} = \int p_0^2/2\rho_{air}c_{air}$, $P_{out} = \int |p|/2\rho_{air}c_{air}$. As shown in Fig. 8b, the red (green) line TL-sim (TL-viscous) represents the simulated TL with (without) thermo-viscous losses taken into account while the blue-dotted line represents the TL obtained from measurements. The shaded areas in both panels represent the bandwidth of the topological edge states. The numerical simulations indicate that thermo-viscous effects are not relevant. However, the comparison with the experimental TL indicates that the measured TL profile is blue-shifted and the bandwidth of the lower band has also increased with respect to that theoretically calculated. The origin of such disagreements is the small discrepancy in the size of the manufactured rods (see section VI in the **Supplementary Materials** for details). In addition, notice that the TL of modes belonging to the second band is larger than that of the first band since the dispersion relation is almost flat and, consequently, the thermo-viscous losses are enhanced due to the extremely low group velocity. Moreover, the higher value of the ASE, indicating a large spreading of the mode, also contributes to the increase of the TL.

In addition, Fig. 8c shows the measurements of the absolute pressure taken at 3060Hz, as an example, in an area at the output of the interface. This 2D map indicates that the sound propagation has obvious directivity when the waves exit the interface. The underlying physical mechanism of this directional propagation is the symmetry features of the eigenmodes defining the edge states, whose pressure patterns are modulated as shown in Figs. 4 and 5. Therefore, when a particular excited edge state arrives at the rear interface, the radiated pressure exhibits a pattern with a principal directional propagation as shown in Fig. 8c. This directional effect inspired us to achieve acoustic focusing by using two topologically protected edge states obtained in nearby interfaces. Since interfaces II and IV have the same dispersion relationship, we design a sandwich structure composed of SC-A and SC-B with the size of 4×6 cells, as shown in Fig. 8d describing the experiment setup. After the wave emits along interfaces II and IV, due to the directionality of the wave, the wavefront of the outgoing wave at the two ports forms a stable crossing area, also called energy convergence, showing the characteristics of acoustic focusing. The measurements shown in Fig. 8e illustrate this property at two different frequencies, where the output beams with principal directions ±30° interfere in the area in front of the sample. For further details about directivity propagation, the reader is addressed to **Appendix C**.

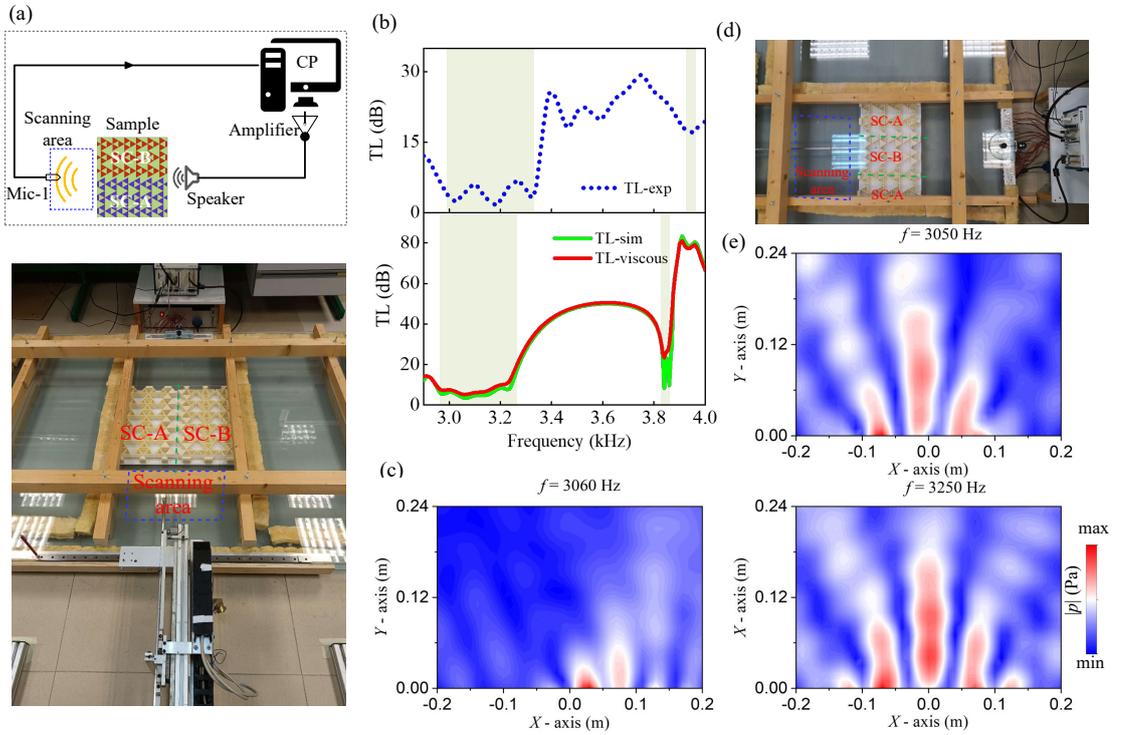

FIG. 8. (a) Experiment setup. Upper panel and bottom panel show the scheme of the data acquisition process and scene photo, respectively. (b) Transmission loss (TL) calculated (bottom panel) and measured (upper panel). The red and green lines define

the TL obtained with and without thermal viscous losses are taken into account. The blue-dotted line represents the TL obtained from measurements. The shadowed areas define the bandwidth of the edge states. (c) The absolute pressure pattern at the scanning area obtained experimentally. (d) Experiment setup for acoustic focusing. (e) Absolute pressure distribution in scanning area obtained experimentally at 3050Hz (top panel) and 3250Hz (bottom panel). Color represents the absolute pressure.

## V. CONCLUSIONS

To summarize, this work demonstrates the unique properties exhibited by topological states related with quasi-type-II DP in acoustic Suzuki phase crystal. Particularly, they are different with the chiral edge states in acoustic quantum Hall effect, the valley vortex edge states in acoustic valley Hall effect and spiral edge states in acoustic quantum spin Hall effect, the edge sates in ASP are characterized by symmetry and directionality. The rich information acquired on topological states in the ASP can be harnessed to achieve interesting acoustic functionalities. For instance, using states belonging to deaf bands to create an acoustic diode, employing Shannon entropy for multi-channel and selective sound transmission, and exploiting directionality to produce acoustic focusing. These features open new paths for the design of highly integrated multifunctional acoustic devices. Let us remark that our results in acoustics can be extended to other classical waves, such as electromagnetic and elastic waves, which can enrich further the study of topological states.

## APPENDIX A: NUMERICAL SIMULATIONS

The numerical simulations are performed using the commercial software, COMSOL Multiphysics, which is based on the finite-element method. We employ the "Pressure Acoustic module" to visualize the modal characteristics and propagating features of acoustic waves. The materials involved in the simulations are the air (the background material) and the ABS (Acrylonitrile Butadiene Styrene) employed in manufacturing the scatterers. The mass density and velocity of air are $\rho_{air} = 1.25 \text{ kg/m}^3$ and $c_{air} = 343 \text{ m}/s$, respectively. The physical parameters for ABS are $\rho_{ABS} = 1060 \text{ kg/m}^3$; $c_{ABS} = 2400 \text{ m}/s$. Due to the large mismatch between the acoustic impedances of the air and the ABS, in the calculations, we consider that the scatterer acts as a rigid body. The mesh type is free triangular, and the largest mesh element size is lower than one-fifteenth of the shortest incident wavelength. In the band structure calculations, Floquet conditions are imposed on the boundaries of the unit cell. For the simulation of wave transmission, the sound source of the structure adopts the background pressure field with a width of 4*a*. Perfect matching layers are imposed around the structure. When calculating the transmission loss, the integral path for the entrance is the boundary between the background pressure field and the structure. The integral path at the exit is parallel to the path of the entrance with the same length 4a, which is located 0.5mm outside the structure. To analyze the effects of viscous and thermal losses caused by scatterers in the structure, we applied the "Thermoviscous Acoustics, Frequency Domain" interface. During the thermo-viscous simulation calculation process, the viscous boundary layer thickness is defined at the maximum value under study. The viscous boundary layer thickness $\varepsilon_{visc}$ is set by $\varepsilon_{visc} = \sqrt{2\mu/(\omega\rho_0)}$, where $\mu$, $\omega$ are dynamic viscosity and angular frequency.

## APPENDIX B: EXPERIMENTS SETUP

The rods of the sample are fabricated using ABS via 3D printing. The sample in Fig. 6a (Fig. 6d), which consisted of 108 (72) triangular rods, was embedded in a 2D planar waveguide formed by two Plexiglas plates. The height of the rods is 30mm. In this case, the 2D approximation is applicable, due to the waveguide supports propagating mode uniformly along the rod-axis for the wavelengths under consideration. Experiments were conducted by a linear array of 16 speakers, to create an incident beam with a Gaussian profile. The microphone is fixed to the coordinate position system and scans an area of 40 cm × 24 cm at the output of the sample. Both the card (NI PXIe-6259) for data acquisition from Mic-1 and the sound card (NI PXI-6723) to regulate the emission waves are integrated into the CP (computer). The scanning frequency interval is 10Hz.

## APPENDIX C: DIRECTIONAL PROPAGATION AND FOCUSING

According to the simulations and experiments, the waves exiting the structure show obvious directionality. The underlying physics explaining such directional behavior comes from the eigenmodes defining the topological states on the interface. When the incident plane wave excites the eigenmodes at the interface, the pressure field is modulated by the eigenmodes, thus forming directional sound propagation.

Figure 9a schematically shows the directional property of the edge states at Interface IV; when waves with a specific range of frequencies propagate along Interface IV to the outside field, the sound waves exhibit a principal propagation direction at a certain angle $\delta$. For example, for the frequency 3060Hz, Fig. 9 shows the experimental 2D map of the absolute pressure measured in the scanned area behind the sample. This measurement confirms the theoretical prediction and simulations. Figure 9c depicts the absolute value of the pressure along different angles $\delta$. The different lines are extracted from the experimental 2D map shown in Fig. 9b. It is observed that the radiated wave has a principal beam along 30º. Small side lobes appear at 60º and 90º. There are also components along $\delta = 150°$ and 120°, but they can be neglected in comparison with the case at the angle 30°.

The frequencies of edge states at interfaces IV and II are the same but the orientation of the triangular scatterers is interchanged along the interface. These features can be exploited to design an acoustic focusing device as schematically explained in Fig. 9d. It is observed how the proposed sandwiched structure is composed of SC-A and SC-B and contains interfaces IV and II. Thus, it is expected that the waves radiating from the interfaces will interfere producing a focal point (red dot) in the radiation field. The simulation results of a 4×4 sandwich structure for the absolute pressure of a wave with a frequency of 3050 Hz are shown in Fig. 9e, where a focal area is clearly shown in the radiation field. In addition, Fig. 9f shows a comparison between the profiles of the radiation field along the line passing through the focal point. The black arrows in Fig. 9e define such a line. Notice that there is a good agreement between the experimental (blue symbols) and simulated (blue continuous line) profiles, supporting the possibility of obtaining focusing devices from topologically protected edge states. In addition to phononic crystals with external lenticular shape [60], gradual material dispersion with gradual geometrically curved surface [61-64], and gradient refractive indices in artificial focusing devices [65-67], using the directivity of edge states can also be an effective way to achieve focusing.

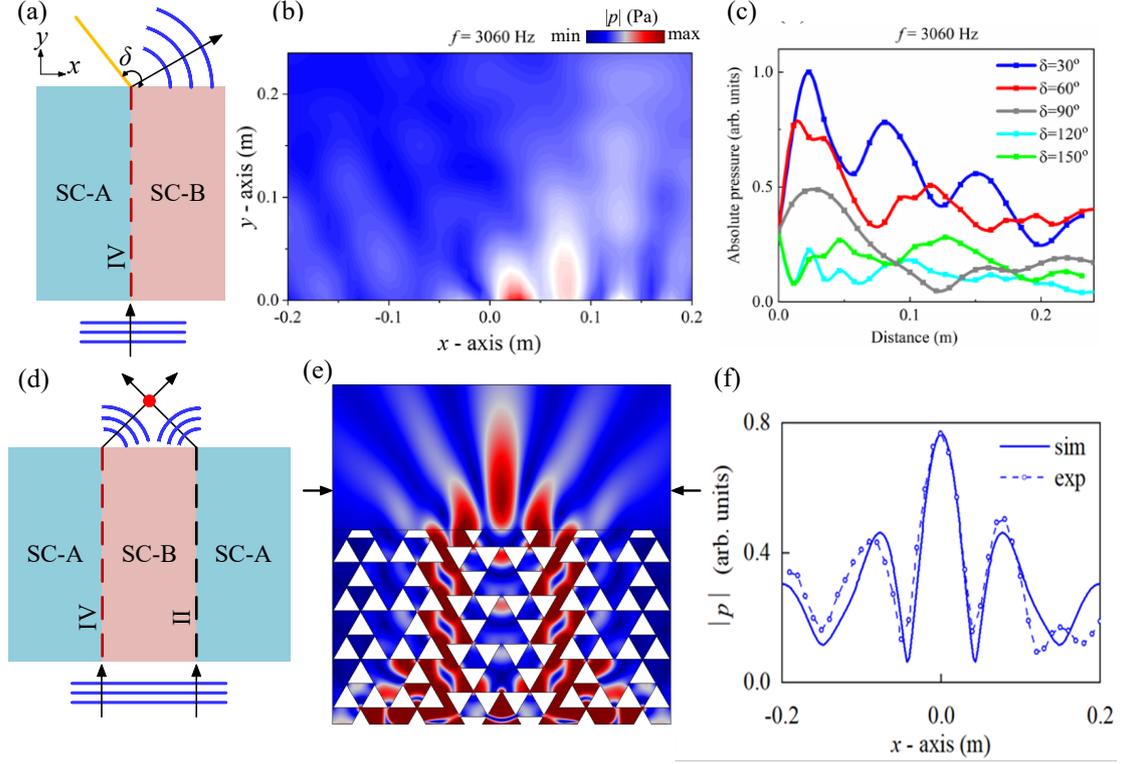

FIG. 9. (a) Schematic diagram representing the propagation of a planar wavefront impinging a structure containing the Interface IV. The horizontal blue lines represent the impinging wavefront; the red dotted line represents the excited edge state while the arrow, which forms an angle of $\delta$ with the *x*-axis, represents the main propagation direction. (b) Snapshot of the absolute pressure map measured in the scanned area using waves of frequency 3060Hz. (c) Absolute pressure measured along different output directions $\delta$. They are obtained from the map in (b). (d) Schematic diagram of the acoustic focusing device containing interfaces II and IV. (e) Numerical simulations showing the focusing produced by an incident plane wave at 3050Hz propagating through the sandwiched structure in (d). The color scale represents the absolute pressure. (f) Profile of the absolute pressure along the line defined by the arrows in (e). The dashed line with symbols and the continuous line represents the experimental and simulated profiles, respectively. For a better comparison, the experimental data are normalized to the peak value found in the simulations.

## ACKNOWLEDGMENTS

This work was supported by the National Natural Science Foundation of China (NSFC) under Grant Nos. 52250287. Zhen Huang acknowledges a scholarship provided by the China Scholarship Council (CSC) under Grant No. 202206280162. The Wave Phenomena Group acknowledges the support of the RDI grant PID2020-112759GB-I00 funded by MCIN/AEI/10.13039/501100011033.

# Supplementary materials for

# Topological transmission in Suzuki phase sonic crystals

Zhen Huang[1,2], Francisco Cervera[1], Jiu Hui Wu[2]*, Martin Ibarias[1], Chongrui Liu[2], Victor M. García-Chocano[1], Fuyin Ma[2]*, José Sánchez-Dehesa[1]*

[1]Wave Phenomena Group, Universitat Politècnica de València, Camino de vera s.n. (Building 7F), ES-46022 Valencia, Spain

[2]School of Mechanical Engineering, Xi'an Jiaotong University, Xi'an 710049, China.

*Corresponding author: ejhwu@xjtu.edu.cn, xjmafuyin@xjtu.edu.cn, jsdehesa@upv.es

# Supplementary materials for

# Topological transmission in Suzuki phase sonic crystals

# I. Dirac point in Suzuki phase and triangular lattice

It is exciting to discover the double degenerate point in the acoustic Suzuki phase (ASP) lattice since double degeneracy is the paramount condition for realizing the topological edge states in acoustic valleys. However, different from the type-I Dirac point (DP) at the corners of FBZ due to the crystal symmetry[1], the accidental DP formed by the 4$^{th}$ and 5$^{th}$ bands in the ASP is located at low-symmetry point in FBZ, and the cone related with the DP lacks obvious tilted features in specific direction[2,3]. Therefore, it can be denominated as a quasi-type-II DP. Next, we analyze the reasons explaining the occurrence of quasi-type-II DP. As shown in Fig. S1a, the sonic crystal under study can be regarded as a rectangular lattice of vacancies created in the well-known triangular lattice. The blue dotted hexagon represents the unit cell of the triangular lattice with the primitive vectors $A_1$ and $A_2$, and $|A_1| = |A_2| = a$. Thus, the complete crystal can be obtained by repetition of the unit cell (containing a single scatterer) at the lattice positions generated from $A_1$ and $A_2$. However, the triangular lattice can be also generated using as primitive vectors $a_1$ and $a_2$, previously used to describe the ASP. The unit cell of the ASP, represented by a blue-dotted rectangular box, contains three scatterers instead of four since the red triangles represent the vacancies created in the original triangular lattice. When $a = 50$mm, $R = 0.45a$, the band structure of the ASP is shown in Fig. S1b, where I$_1$ defines the position between the 4$^{th}$ and 5$^{th}$ bands where the quasi-type-II DP appears. The inset represents the unit cell that is repeated at the lattice position generated $a_1$ and $a_2$. For the sake of comparison, Figs. S1c and S1d represent the band structures of the triangular lattice calculated with the larger unit cell generated by the rectangular lattice and with the standard unit cell. Note that the frequency positions, I$_2$ and I$_3$, appear at the same frequency in both figures. In fact, due to the scatterer displaying $C_{3v}$ symmetry consistent with the triangular lattice, I$_3$ located at the corner of the FBZ, is a type-I Dirac point[4]. The frequencies of points I$_1$, I$_2$ and I$_3$ in the three band structures are the same, which indicates that there should be a connection between them.

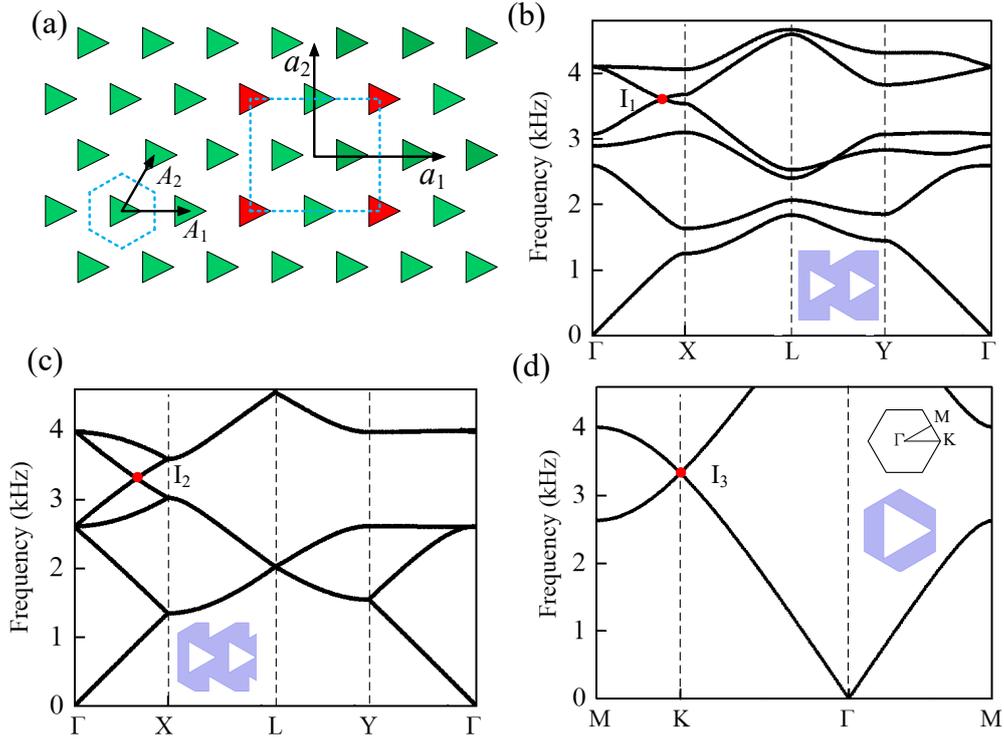

FIG. S1. (a) Schematic diagram of a sonic crystal consisting of scatterers with triangular section distributed in a triangular lattice. The blue dashed hexagon represents the primitive unit cell of the triangular lattice defined by the vectors $A_1$ and $A_2$. The rectangle represents the unit cell of the acoustic Suzuki phase (ASP) in which the unit cell contains three scatterers. The red triangles represent the vacancies created in the triangular lattice. (b) Acoustic band structure of the ASP. The inset shows the unit cell of the ASP. The DP is denominated as $I_1$. (c) The dispersion structure of the lattice represented by the blue dotted rectangular box in (a). The 4$^{th}$ and 5$^{th}$ bands appear a Dirac cone point $I_2$ in the direction of ΓX. The inset represents the cell structure. (d) Band structure of the triangular lattice obtained with the unit cell shown in the inset. A type-I Dirac point $I_3$ appears at the corner point K.

## II. Berry curvature in acoustic Suzuki phase

Topological phase transitions have been marked one of the most important highlights in recent condensed matter physics. In the frequency band structure, for an acoustic system with energy valleys due to band degeneracy, an important method to measure the topological non-trivial phase is to calculate the valley-Chern number. The Chern number can be obtained by integrating the Berry curvature $F_n(k)$ over the wave vector $k$-space in a two-dimensional (2D) torus[5]. For example, for the $n$th band in the band structure, the formula to calculate its Chern number $C_n$ in the continuous space is defined as [6,7],

$$C_n = \frac{1}{2\pi} \iint F_n(k) d^2k. \tag{1}$$

In order to simplify the calculation, the result of the Chern number can be obtained on a discretized Brillouin zone[6]. Here, we use the general approach, which discretize the two boundary vectors $k_{j_s}$ ($s$=1, 2) of the first Brillouin zone with a rectangle shape into $N_1$ and $N_2$ points respectively. Then in the 2D vector space, the coordinates of any point are denoted as

$$k_b = (k_{j_1}, k_{j_2}) \quad (b=1, 2, \ldots, N_1N_2), \tag{2}$$

where $k_{j_s} = \frac{2\pi}{a_s} \cdot \frac{j_s}{N_s}$ ($j_s$=0, 1, …, $N_s$-1).

Setting $N_s = a_\mu N_B$ ($s \neq \mu$), in this way, the unit piece is a square with the size $2\pi/(a_1 a_2 N_B)$. For each discrete unit piece with area $\Delta S_k$, the integral value of Berry curvature on it is

$$F_n(k)\Delta S_k = \frac{1}{2\pi} \text{Im}(ln[U_{k_1 \to k_2}(k_b) U_{k_2 \to k_3}(k_b) U_{k_3 \to k_4}(k_b) U_{k_4 \to k_1}(k_b)]. \tag{3}$$

Here, $U_{k_h \to k_g}(k_b) = \left| \left\langle \Psi_{n,k_{j_h}}(k_b) \middle| \Psi_{n,k_{j_g}}(k_b) \right\rangle \right|$, $h, g$=1,2,3,4, and the inner product in the formula can be defined as

$$\left\langle \Psi_{n,k_{j_h}}(k_b) \middle| \Psi_{n,k_{j_g}}(k_b) \right\rangle = \iint \Psi^*_{n,k_{j_h}}(r) \cdot \widehat{\mathcal{B}}(r) \cdot \Psi_{n,k_{j_g}}(r) dS, \tag{4}$$

where the $\Psi_{n,k}(r)$ is the $k$-dependent wavefunction of the $n$th band in the acoustic wave; $\widehat{\mathcal{B}}(r)$ is the energy density operator, and its function is to make the integral on the right side of the Eq. (4) proportional to the energy of the sound wave when $k = k'$. For acoustic waves, the energy density operator is

$$\widehat{\mathcal{B}}(r) = \frac{1}{2\rho(r)c(r)^2}, \tag{5}$$

where $\rho$ is the mass density, and $c$ is the speed of sound. Both the density and speed depend on the medium at the position *r* [8].

Note that for an acoustic Bloch wavefunction $\Psi_{n,k}(r) = e^{ikr}p_{n,k}(r)$, during the calculation, the periodic part of the Bloch function, $p_{n,k}(r)$, which gives the acoustic pressure distribution for the *n*th acoustic band with wave-vector *k*, needs to be normalized. The formula is defined as [8]

$$\mathcal{A}_{n,k} = \iint p_{n,k}^*(r) \cdot \frac{1}{2\rho(r)c(r)^2} \cdot p_{n,k}(r) dS. \tag{6}$$

Therefore, a new inner product expression needs to be defined on the basis of formula (4), as follows

$$\left\langle \Psi_{n,k_{j_h}}(k_b) \middle| \Psi_{n,k_{j_g}}(k_b) \right\rangle = (\mathcal{A}_{n,k_{j_h}} \mathcal{A}_{n,k_{j_g}})^{-1/2} \iint \Psi_{n,k_{j_h}}^*(r) \cdot \widehat{\mathcal{B}}(r) \cdot \Psi_{n,k_{j_g}}(r) dS. \tag{7}$$

The integral of the Berry curvature on each discrete unit block can be obtained from Eqs. (3) and (7). Then, the Chern number of Eq. (1) can be calculated by adding up the integral values on all discrete unit pieces, and the expression is given by

$$C_n = \sum_{b=1}^{N_1 N_2} F_n(k) \Delta S_k. \tag{8}$$

According to Eq. (8), based on the distribution of $F_n(k)\Delta S_k$ in parameter space, that is, the Berry curvature, the topological properties of the energy band can also be characterized. The calculation results are plotted in Figs. 4e and 4f in the main content.

## III. Band structure of topological trivial insulator

In the process of scatterer rotation, the SCs before and after degeneration of DPs have different topological phases, and there are topologically protected edge states of the interface composed of two SCs with distinct topological phases[9]. While there are no edge states on the interface separating two SCs with the same topological phase. As shown in Fig. S2a, from the band structure of ASP with the rotation angle *β*=25° (blue vertical short line) and *β*=35° (the red horizontal short line), the two cases are nearly the same. At *k*=0.6695, eigenstates M1 and M2 are taken from the fourth and fifth bands of a SC with *β*=35° (SC-C), and M3 and M4 from the fourth and fifth bands of a SC with *β*=25° (SC-D). Figure S2b depicts the pressure patterns of these four eigenmodes, showing that the modes of the two SCs are the same, and therefore the two SCs have the same topological phases.

Figure S2c schematically depicts a vertical supercell containing Interface T-I composed of SC-C (upper) and SC-D (lower), and Interface T-II composed of SC-C (lower) and SC-D (upper). The supercell contains 40×1 unit cells, of which both ends contain 10×1 unit cells. The red dashed rectangle shows zoomed views of the two interfaces. Fig. S2d shows the projected band structure of the vertical supercell, in which there is no edge states within the overlapping bulk bandgap. Moreover, the same situation exists in the projected bands (Fig. S2f) of the horizontal supercell (Fig. S2e). It can be seen from Fig. S2 that there are no topological edge states on the interface composed of SCs with the same topological phase.

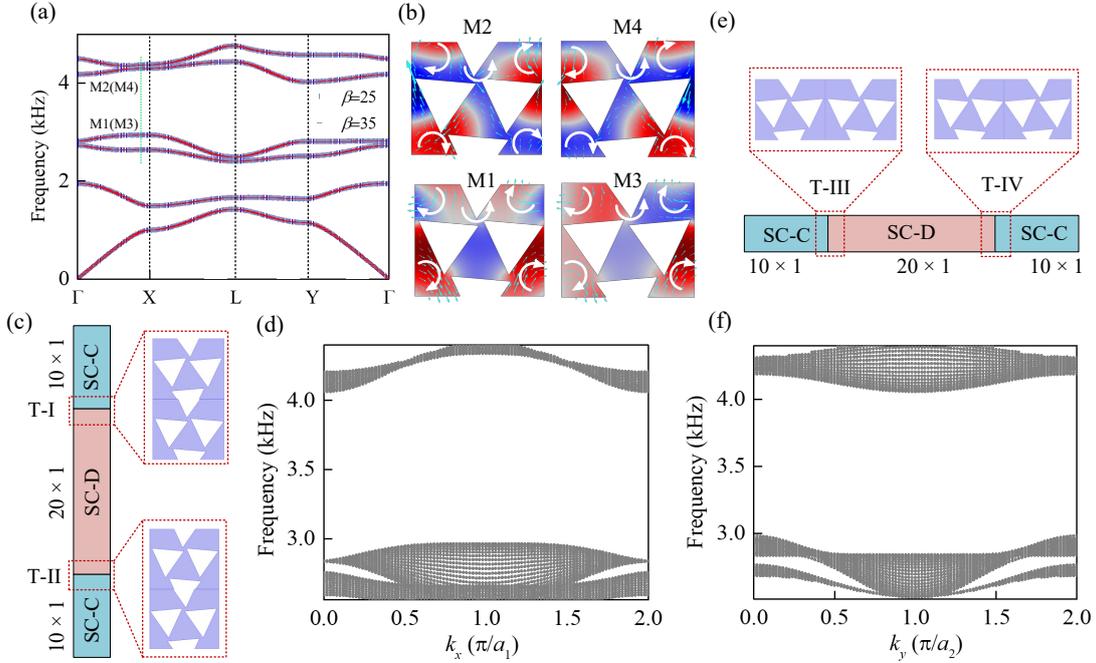

FIG. S2. (a) Band structure with different rotation angles. The blue vertical short line and the red horizontal short line represent the band structure when the rotation angle $\beta=25°$ and $\beta=35°$, respectively. The green dotted line indicates $k=0.6695$, and the intersection of SC with the fourth and fifth bands at an angle of rotation of 25° (35°) is indicated by M1 (M3) and M2 (M4). (b) Sound pressure mode of the states M1-M4. The color represents the sound pressure (real part), in which red and blue colors represent the positive and negative values, respectively. The cyan arrow indicates the sound intensity, and the white circular arrow indicates the direction of the sound energy

flow. (c) Scheme of the superlattice composed of SC-C and SC-D with interfaces T-I and T-II. The red-dotted rectangular box deploys a zoomed view of the interfaces T-I and T-II. (d) Projected band structure of the supercell (c) around the double degenerate point. The black dots represent bulk bands. (e) Scheme of the superlattice composed of SC-C and SC-D with interfaces T-III and T-IV. The red dotted rectangular box deploys a zoomed view of the interfaces T-III and T-IV. (f) Band structure of the supercell (e) around the double degenerate point.

## IV. Eigenmode symmetry and transmission characteristics

### A. The eigenmodes and acoustic transport along the Interface I

There are three frequency bands, ES-I-1, ES-I-2 and ES-I-3 composed of topological states on the Interface I, and the eigenmodes of the edge states above them have obvious symmetry. For the convenience of expression, we define the sound pressure mode as "1" for the symmetry of the coordinate axis, and "0" for the anti-symmetry of the coordinate axis. Therefore, the symmetry of the sound pressure mode for the third and fourth edge state bands at the higher frequency is the same, and it is marked as "11". The first and second numbers indicate symmetry about the $x$-axis and $y$-axis, respectively. The symmetry of the sound pressure mode in the first and second edge state bands includes "00" and "01". Different symmetries mean that there are also differences in the transport characteristics of the edge states. We consider three plane waves with different frequencies incident in three directions into the acoustic field domain containing Interface I (indicated by the green dotted line), as shown in Fig. S3, the incident frequencies of the first, second and third rows are 2978.3 Hz, 3535.7Hz and 4031.4Hz, respectively. The frequencies in the first and second lines are the frequencies of $A_2$ ($A_2'$) and $B_2$ ($B_2'$) when $k$=0.5 (1.5) in the band ES-I-1 (ES-I-2), and the third line is the frequency of E when $k$=0.1 in the band ES-I-3. Each column in the figure represents incident waves in the same direction. In the first column, wave direction is [1, 0, 0], which shows that the transmission of wave with the frequency of two edge states (with sound pressure mode symmetry of "00" or "01") is suppressed, while the higher frequency (acoustic pressure mode symmetry of the edge states is "11") can achieve acoustic transmission. For the case where the sound pressure symmetry is "00" or "01", the direction of the sound pressure modes is perpendicular to the incident direction, so it cannot be excited by this plane wave. Therefore, relative to the incident direction [1, 0, 0], the frequency band formed by the edge states with lower frequency cannot be excited. However, for the edge state with higher frequency, due to the symmetry about the $x$-axis, this means that its eigenmode has a similar symmetry to be excited by the plane wave incident along the $x$-axis direction. From the second and third columns, when the wave vector direction is along the $y$-axis, including wave direction [0, 1, 0] and [0, -1, 0], and incident waves of different frequencies can propagate along the interface. This is because the acoustic pressure modes of the lower-frequency edge states are antisymmetric about the $x$-axis, that is, the normal direction of the isophase plane is parallel to the $y$-axis, so the plane waves propagating along the $y$-axis can excite these modes, thereby realize the propagation of sound waves. It is worth noting that for a plane wave with a frequency of 4031.4 Hz, the wave vector can propagate in the sound field no matter whether it is along the $y$-axis or along the $x$-axis, but the propagation loss is different.

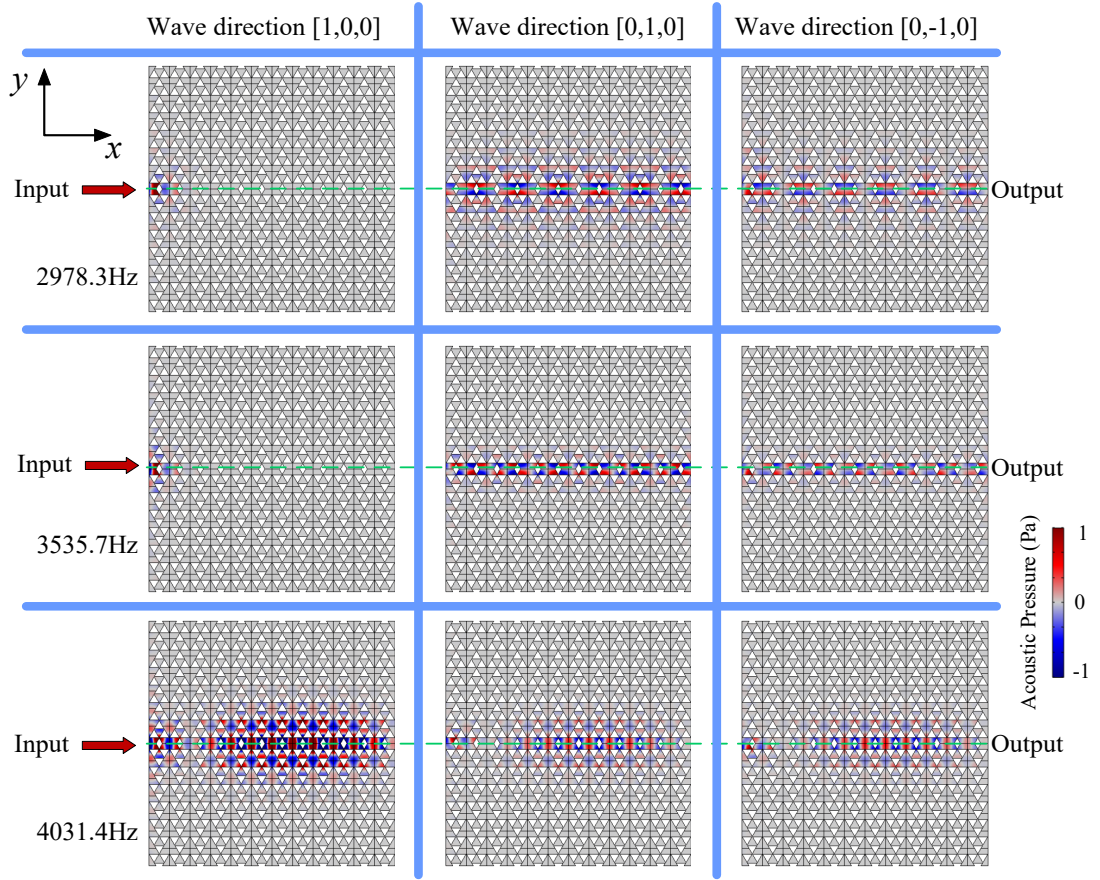

FIG. S3. Transmission of incident waves with different wave vector directions incident on the sound field domain containing Interface I at the frequencies of edge states $A_2$ ($A_2'$), $B_2$ ($B_2'$) and E

## B. The eigenstates on the Interface III and acoustic wave transmission

The edge states on Interface III have the same symmetry, denoted by "11" (symmetric respect to both *XZ* and *YZ* planes). The sound propagation with wave vectors incident along three different directions is shown in Figure S4, where the green dotted line indicates Interface III. The frequencies correspond to edge states with $k = 0.5$ (1.5), corresponding to modes $A_1$ ($A_1'$) and $B_1$ ($B_1'$). It is observed that waves are transmitted along Interface III, but their transmission loss is different for the different impinging directions. This result can be understood by looking at the pressure patterns of the edge modes shown in Fig. 5 of the main text, where the sound pressure modes on Interface III are represented. It can be shown that the 0Pa sound pressure contour line on the interface is not parallel to the coordinate axis, so the normal line of the sound pressure isophase plane has components in both *x*-axis and *y*-axis, and the two quantities are different.

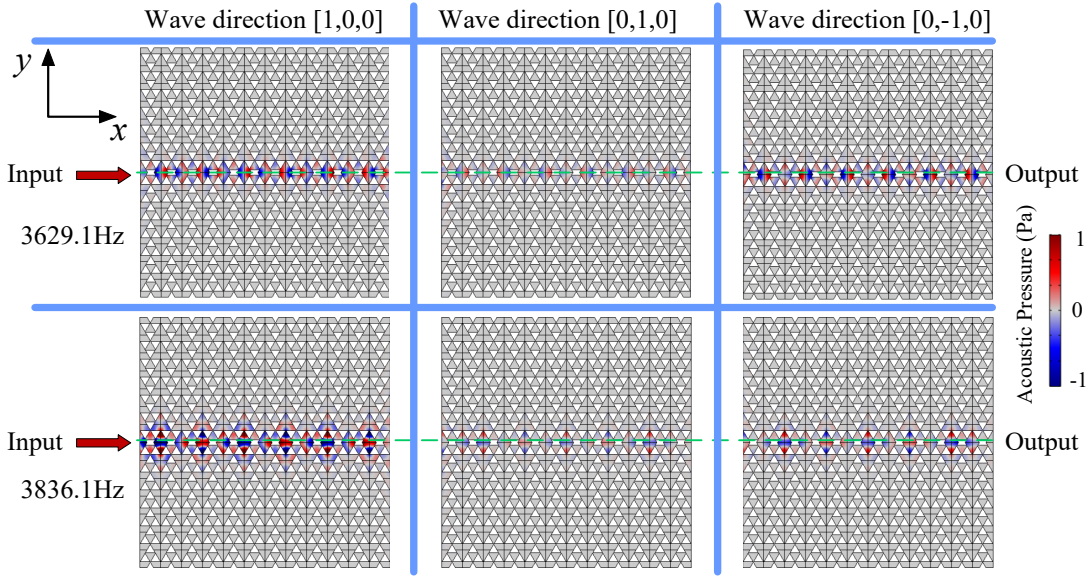

FIG. S4. Sound transmission along Interface III with impinging waves coming from different directions with frequencies corresponding to edge states $A_1$ ($A_1'$) and $B_1$ ($B_1'$).

## C. Edge states localized at Interface II (IV) and acoustic wave transmission

For the case of edge states associated with interfaces II and IV, they do not exhibit any defined symmetry about the *XZ* or *YZ* planes and, therefore, the propagation of impinging waves along different directions cannot be predicted by analyzing the symmetry of the eigenmodes. Figure S5 shows the transmission of incident plane waves with frequencies 2985.9 Hz and 3866.7 Hz. The incident directions from the first row to the third row are [-1, 0, 0], [1, 0, 0], and [0, 1, 0] respectively. The first two columns represent the propagation of sound waves along Interface II, while the last two columns represent the propagation of sound waves cross Interface IV. In an acoustic field containing different interfaces, when the frequency of the incident wave is higher, the transmission of the acoustic wave on the interface is hardly affected by the direction of the incident wave. However, for the frequencies of lower edge states, the acoustic wave transmission exhibits a correlation with the incident wave direction. For example, for an incident wave with a frequency of 2985.9 Hz, when it propagates in the acoustic field domain including Interface II, the transmission of the wave direction [1, 0, 0] is suppressed. In the acoustic field domain containing Interface IV, the transmission of the wave direction [-1, 0, 0] is suppressed. This phenomenon can be used to design directional topological acoustic emission.

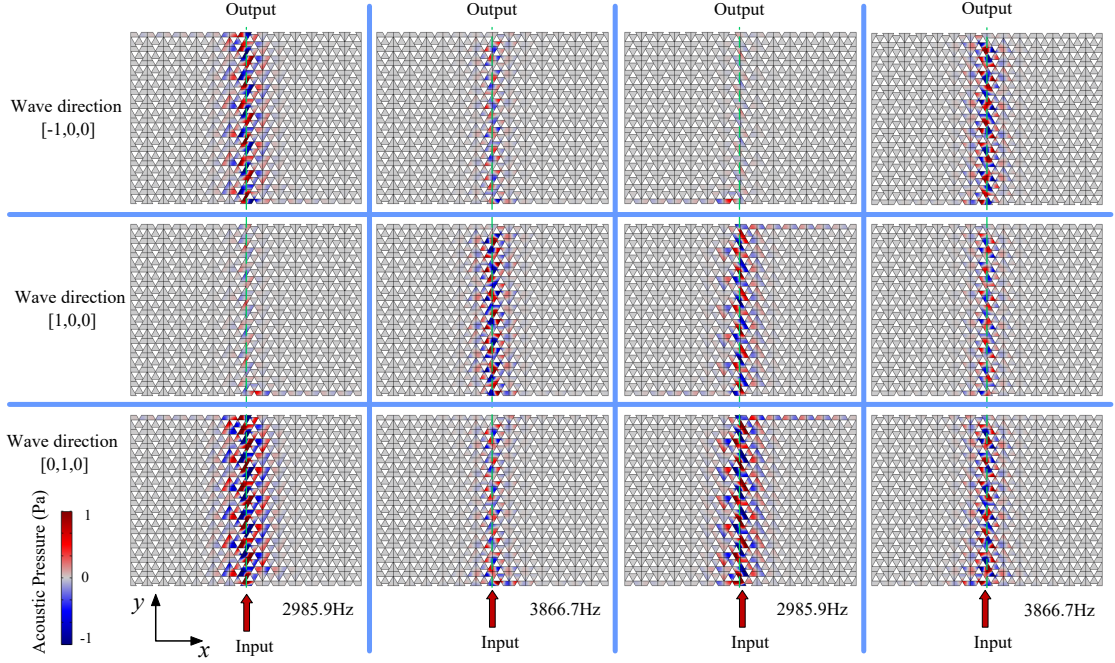

FIG. S5. Snapshots of the sound transmission along Interface II (IV) produced by plane waves with different incident directions. The frequencies correspond to edge states $A_3$ ($A_4$) and $B_3$ ($B_4$).

**D. Deaf bands and transmission**

From the pressure field patterns of edge modes in Fig. 5 (main text), we observe that each mode exhibits different symmetry. Particularly, the modes in bands ES-I-1 and ES-I-2 have the planes of equal phase along the perpendicular $y$-direction. Therefore, these modes cannot be excited by impinging waves traveling along the [1, 0, 0] as shown in Fig. S3. Therefore, the modes in bands ES-I-1 and ES-I-2 can be considered as deaf acoustic modes[10] for waves propagating along the [1, 0, 0] direction.

It should be noted that when the incident wave is not symmetric with respect to the interface, the incident wave can generate additional non-negligible components in the transversal direction to the interface, allowing the excitation of deaf edge modes. This effect is shown in Fig. S6 where it is observed the excitation of an edge states with an impinging wave along the direction [1, 0, 0]. Figure S6a show the configuration when the excitation source is symmetric with respect to the interface. With this configuration, Fig. S6b shows that the edge mode with frequency of 3600 Hz cannot be excited since its symmetry is orthogonal with respect to the symmetry of the impinging beam. However, when the exciting beam is shifted up, when entering the acoustic field, an effective component perpendicular to the initial incident wave vector is generated, so that the incident wave in the deaf band in the $x$-axis direction can move along the Interface I, as shown in Fig. S6d. When the incident plane wave (the line source is used in the simulation, the domain source here is a schematic diagram) is symmetrical about the interface. In other words, when the interface passes through the center of the incident line source, the incident field does not generate the effective component perpendicular to the vector of initial incident wave, before propagating along the interface. So, the incident wave is localized at the entrance. When the interface deviates from the center of the line source, an effective component perpendicular to the interface is generated in the incident acoustic field, so the intrinsic acoustic pressure mode on the Interface I is excited, and the topologically protected transmission of the edge state is realized. Therefore, the transmission through edge states belonging to deaf bands can be realized by changing the position of the incident sound field, thus turning the deaf modes into transmitted modes.

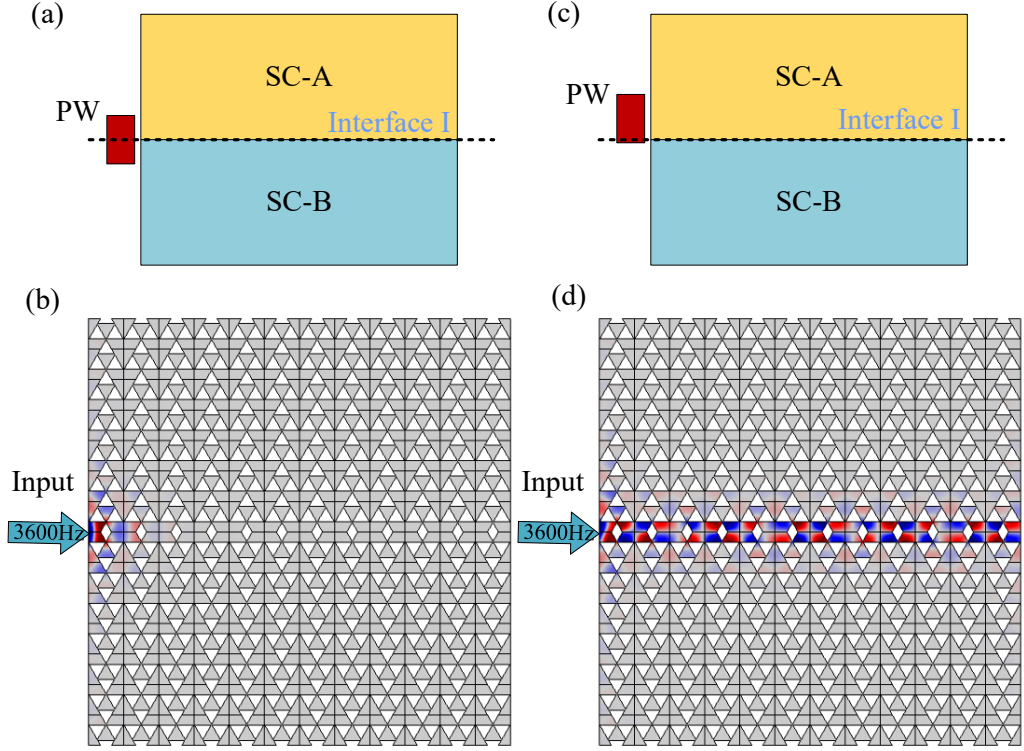

FIG. S6. (a) Scheme of the structure defining Interface I where the exciting beam (PW) is symmetric with respect to the interface. (b) Snapshot of the total pressure pattern obtained with the previous configuration when the incident beam has frequency 3600 Hz (c) Scheme of the structure defining Interface I where the exciting beam (PW) is non-symmetric with respect to the interface. (d) Snapshot of the total pressure obtained with the previous configuration at the frequency of 3600 Hz. Now the deaf edge mode has been excited due to the asymmetry of the impinging beam.

## V. Transport along interfaces containing topological edge states

The paramount property of topological non-trivial systems is the exhibition of unidirectional and robust propagation with edge states existing at the interfaces between distinct topological phase zones. As described above, the symmetry of the pressure patterns of the edge states enriches the transmission features through topological edge modes. On the other hand, some asymmetric scattering, producing an effective component perpendicular to the incident wave vector, can easily modify the incident exciting plane wave. Therefore, when 2D plane waves impinge the structure through the four different interfaces, a variety of behaviors in transmission appear depending on the incident directions.

The chart shown in Fig. S7 reports the transmission properties of plane waves impinging the different ports, where letters "Y" and "N" indicate that transmission is allowed and not, respectively. The number "0" indicates that no edge state with such frequency exists at the corresponding interface. Therefore, the chart presents a comprehensive study of the transmission properties along the interfaces.

| | EG-A | | | EG-C | | | EG-E | | | EG-F | | |
|---|---|---|---|---|---|---|---|---|---|---|---|---|
| Port-I = Input | Port-II | Port-III | Port-IV | Port-II | Port-III | Port-IV | Port-II | Port-III | Port-IV | Port-II | Port-III | Port-IV |
| [1, 0,0] | N | N | N | N | N | N | 0 | 0 | 0 | 0 | 0 | 0 |
| [0, 1,0] | Y | N | Y | N | Y | N | 0 | 0 | 0 | 0 | 0 | 0 |
| [0, -1,0] | Y | N | Y | N | Y | N | 0 | 0 | 0 | 0 | 0 | 0 |
| | EG-A | | | EG-C | | | EG-E | | | EG-F | | |
| Port-II = Input | Port-I | Port-III | Port-IV | Port-I | Port-III | Port-IV | Port-I | Port-III | Port-IV | Port-I | Port-III | Port-IV |
| [0, 1,0] | Y | N | Y | 0 | 0 | 0 | N | Y | Y | N | N | Y |
| [1, 0,0] | Y | N | Y | 0 | 0 | 0 | N | Y | Y | N | N | Y |
| [-1, 0,0] | Y | N | Y | 0 | 0 | 0 | N | Y | Y | N | N | Y |
| | EG-A | | | EG-C | | | EG-E | | | EG-F | | |
| Port-III = Input | Port-I | Port-II | Port-IV | Port-I | Port-II | Port-IV | Port-I | Port-II | Port-IV | Port-I | Port-II | Port-IV |
| [-1, 0,0] | 0 | 0 | 0 | Y | N | N | N | Y | Y | 0 | 0 | 0 |
| [0, 1,0] | 0 | 0 | 0 | Y | N | N | N | Y | Y | 0 | 0 | 0 |
| [0, -1,0] | 0 | 0 | 0 | Y | N | N | N | Y | Y | 0 | 0 | 0 |

FIG. S7. Chart of the transmission properties, where letters "Y" and "N" indicate if the transmission is allowed or not, respectively. The value "0" means no propagation.

## VI. Frequency shift of edge states due to rod size

The sample fabricated by the 3D printer shrinks to a certain extent after cooling, so the actual size of the rods was slightly smaller than that resulting from the design process ($R = 0.54a$), which has led to small differences between the numerical simulations and the experimental results. For example, $R = 0.51a$ is considered here in Fig. S8a, and the dispersion relation of the supercell containing Interface II is shown in Fig. S8b, where the blue and orange lines define the frequency bands containing topological states. These two bands have bandwidths similar to those obtained experimentally, as indicated by the shaded areas in Fig. S8c. Therefore, we concluded that the frequency blue-shift observed experimentally in Fig. 8b is attributed to the rod dimension, which is slightly smaller than that obtained in the design process.

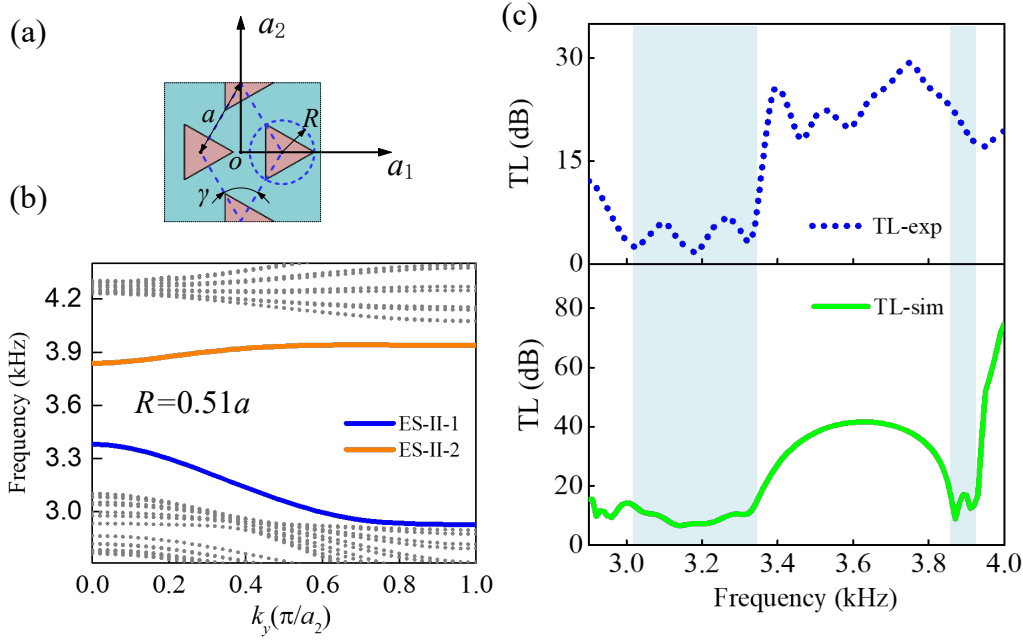

FIG. S8. (a) Scheme of the unit cell. The distance between rods is $a$, and the radius of the outer circle of the triangular rods is $R$. (b) The dispersion relation of the supercell containing Interface II with $R = 0.51a$. The gray dots represent bulk modes. The blue and orange lines represent the bands ES-II-1 and ES-II-2 composed of topological states, respectively. (c) Transmission loss (TL) obtained from experiments (upper panel) and numerical calculations (bottom panel). The green line represents the numerical simulations in the inviscid approximation. Shaded regions define the bandwidth of bands ES-II-1 and ES-II-2.